\documentclass[lettersize,journal]{IEEEtran}
\usepackage{amsmath,amsfonts}
\usepackage{algorithm}
\usepackage{array}
\usepackage[caption=false,font=normalsize,labelfont=sf,textfont=sf]{subfig}
\usepackage{textcomp}
\usepackage{stfloats}
\usepackage{url}
\usepackage{verbatim}
\usepackage{graphicx}
\usepackage{cite}

\usepackage{multirow}
\usepackage{booktabs}

\usepackage{algorithm}
\usepackage{algpseudocode}

\usepackage{algorithmicx}
\usepackage{amssymb}

\usepackage{amsthm}
\theoremstyle{definition}
\newtheorem{definition}{Definition}

\newcommand{\ConcurGuard}{\textsf{ConFixAgent}}

\newcommand\lnote[1]{#1}
\newcommand{\PFIX}{\textsf{PFIX}}
\newcommand{\HIPPODROME}{\textsf{HIPPODROME}}

\usepackage{tcolorbox}

\newtcolorbox{resultbox}{
  colback=gray,
  colframe=gray,
  rounded corners,
  boxrule=1pt,
  arc=5pt,
  left=6pt,
  right=6pt,
  top=6pt,
  bottom=6pt,
  boxsep=0pt,
  beforeafter skip=10pt
}

\begin{document}

\title{An End-to-End Approach for Fixing Concurrency Bugs via SHB-Based Context Extractor}

\author{Zhuang Li, Qiuping Yi, Keyang Xiao, Zongcheng Ji, Hongliang Liang
\thanks{This paper was produced by the IEEE Publication Technology Group. They are in Piscataway, NJ.}
\thanks{Manuscript received April 19, 2021; revised August 16, 2021.}}

\markboth{Journal of \LaTeX\ Class Files,~Vol.~14, No.~8, August~2021}%
{Shell \MakeLowercase{\textit{et al.}}: A Sample Article Using IEEEtran.cls for IEEE Journals}

\IEEEpubid{}

\maketitle

\begin{abstract}
With the rise of multi-core processors and distributed systems, concurrent programming has become essential yet challenging, primarily due to the non-deterministic nature of thread execution. Manually addressing concurrency bugs is time-consuming and error-prone. Automated Program Repair techniques provide a promising solution. However, developing an end-to-end concurrency bug repair tool is particularly challenging. Most existing tools rely on the assumption that bug-related information is readily available or that concurrency bug contexts are ideally extracted, which is often impractical in real-world scenarios.
This paper introduces \ConcurGuard, an LLM-driven agent capable of fixing various types of concurrency bugs in an end-to-end manner, eliminating the need for any prior bug-related information.
Specifically, we propose a novel context extraction approach designed for concurrency bug repair, utilizing Static Happens-Before Graphs to identify bug-relevant sections. We implemented \ConcurGuard\ and evaluated it across multiple benchmark sets. Our extensive experiments demonstrate that \ConcurGuard\ significantly outperforms state-of-the-art tools in addressing diverse types of concurrency bugs, with its context extraction method markedly enhancing the accuracy of LLM-generated repair solutions.
\end{abstract}

\section{Introduction}
\label{introduction}


With the rise of multi-core processors and distributed systems, concurrent programming has become essential in modern software development~\cite{multi-core, bug-survey}. However, writing concurrent code is challenging due to the non-deterministic nature of thread execution, making concurrency bugs common in real-world applications and difficult to detect and resolve~\cite{change, bug-widespread, bug-pattern, bug-survey}.

Studies have shown that the software industry spends over \$100 billion on 
failure diagnosis and repair~\cite{economic}, and programmers spend more than 50\% of 
their time resolving bugs~\cite{time}. Manual bug fixing is not only time-consuming~\cite{bug-survey} but also prone to errors~\cite{swan, bug-survey, hipp}.
To address these issues, Automated Program Repair (APR) 
techniques have been developed to streamline the repair process, 
minimize manual intervention, and ensure higher software quality~\cite{APR}. 
While various methods for fixing concurrency bugs already 
exist~\cite{CaiFix, HuangFix, CFix, DFix, Grail, Axis}, 
our investigation into APR for concurrency bugs 
has revealed three key challenges:



First, gathering concurrency bug information is inherently challenging. In practice, some tools~\cite{CFix, CaiFix, Axis} focus exclusively on repair techniques. They do not address the prior problems of bug detection and localization\footnote{Bug detection refers to identifying deviations from expected behavior—i.e., determining whether a bug exists (often via crashes, or runtime exceptions). Bug localization refers to precisely identifying the code locations responsible for faulty behavior.}—a primary challenge in concurrency bug handling, as noted in prior studies~\cite{swan,fix-survey}.
Other tools employ various detection methods to gather bug information, including run-time analysis, bounded model checking, or static analysis~\cite{HuangFix, hipp, ConcBugAssist, pfix, dyn-location}.
However, these methods frequently encounter effectiveness limitations.
For instance, HIPPODROME, a state-of-the-art static analysis-based tool, establishes correctness criteria focused on resolving data races detected by RacerD~\cite{hipp}. Nevertheless, these criteria suffer from significant limitations: empirical studies show most data races are benign and do not require fixes~\cite{race-harm1,race-harm2}, and concurrency bugs frequently stem from other issues beyond data races, particularly atomicity violations~\cite{oracle,oracle2,bug-survey}.
In contrast, dynamic analysis approaches align better with developers’ intentions, as they ground bug detection in actual runtime behaviors~\cite{oracle,oracle2}. However, they face challenges in accurately localizing concurrency bugs under dynamic conditions. For example, \PFIX—a state-of-the-art dynamic analysis-based tool~\cite{pfix}—struggles with precisely identifying the root causes of concurrency bugs. Due to the complexity of the thread interleaving space, it remains difficult to pinpoint the exact factors that trigger such bugs~\cite{failture-execution}.
Second, existing tools typically specialize in repairing 
specific types of concurrency bugs, 
such as atomicity violations~\cite{CFix, CaiFix}, 
deadlocks~\cite{DFix1, DFix2}, or data races~\cite{hipp}.
Even PFIX~\cite{pfix} —
while effective at addressing all types of concurrency bugs caused by memory interleavings, remains unable to handle deadlocks.
Third, accurately fixing concurrency bugs is a 
significant challenge. Solutions must address bugs 
across all potential scheduling scenarios without introducing new issues. 
This requires comprehensive global reasoning within 
the entire program, which is a complex and difficult task~\cite{pfix,hipp}.

The rise of Large Language Models (LLMs) and their advanced code comprehension capabilities has led to their widespread adoption in various software development tasks, including APR~\cite{LLM-FIX, LLM-FIX1, LLM-FIX2}. However, existing LLM-based APR approaches~\cite{prompt-fix,fine-fix,mt-fix,seq-fix,code-fix,dl-fix,VulRepair,Tfix,llm-patch,cofee-patch,retrive-patch,LLM-FIX2,LLM-FIX1,zero-fix,Dlfix,Inferfix,RepairLLaMA} still cannot fully automate the entire workflow of concurrency bug detection, localization, and repair on raw code as traditional tools like \PFIX~\cite{pfix} and \HIPPODROME~\cite{hipp} do. 
They require active developer involvement: developers must manually identify buggy code segments, annotate lines needing fixes, and even explicitly state the known causes of the bugs~\cite{fine-fix,prompt-fix,llm-patch}. 
This indicates that such approaches generally operate under the assumption that relevant bug information is already at hand, allowing them to concentrate predominantly on patch generation while overlooking the crucial preceding step of acquiring that information.
In practice, however, such information is often unavailable~\cite{swan,bug-survey,hipp}, 
making the acquisition of bug-related information a critical challenge in concurrency bug repair~\cite{swan}.

In this paper, we present \ConcurGuard, an end-to-end LLM-driven automated repair agent designed to address various types of concurrency bugs in Java programs without relying on manually provided bug information. 
\lnote{Its scope comprehensively covers all non-deadlock concurrency bugs—including data races, order violations, and atomicity violations~\cite{pattern}—as well as deadlock concurrency bugs, including both resource deadlocks, which arise from cyclic competition for locks or other tangible resources, 
and 
communication deadlocks, which arise from threads waiting for undelivered intangible signals (e.g., \(wait()/notify()\) on condition variables) and are unrelated to lock acquisition order~\cite{deadlock1}.}

The process begins with the bug detector generating diverse thread interleavings to expose concurrency bugs. 
Once a bug is found, bug reports together with code snippets extracted by the context extractor are sent to the LLM for localization and repair.
This iterative process continues until no bugs are detected or the maximum iteration limit is reached. A key challenge for \ConcurGuard\ lies in designing effective context extraction methods. Unlike non-concurrent programs, where extraction focuses on code near the bug, concurrency bugs stem from interleaved executions, necessitating a global perspective for accurate localization and repair. 
The most similar recent work, Inferfix~\cite{Inferfix}, utilizes Infer~\cite{infer} to extract bug-related code. However, its approach is limited to interleavings between two shared memory accesses.

To achieve effective context code extraction, \ConcurGuard\ leverages a Static Happens-Before Graph to calculate potential triggers of concurrency bugs while preserving concurrent semantics to enhance LLM comprehension. Its context code extractor efficiently extracts fragments that improve the LLM’s understanding and repair capabilities. In evaluations against state-of-the-art tools, \ConcurGuard\ demonstrated superior performance in repairing a wide range of concurrency bugs.

We summarize our contribution as follows:
 \begin{itemize}
 
 \item \textbf{An End-to-End Universal Fixer for Concurrency Bugs}. 
We present \ConcurGuard\ 
as the most comprehensive universal tool available for automatically fixing 
various types of concurrency bugs without requiring any manually provided bug information.

 
 \item \textbf{Automatic Context Code Extraction}. 
\ConcurGuard\ introduces an innovative method for identifying potential triggers of concurrency bugs through a Static Happen-Before Graph. This approach filters over 90\% of tokens and enhances repair accuracy by 14\%. By preserving events that may trigger concurrency bugs while maintaining concurrent semantics, this method effectively reduces input text length, minimizing distractions for LLMs.

\item \textbf{Evidence of Effectiveness}. 
We implemented \ConcurGuard\ and evaluated it on several 
set of benchmarks. Our evaluation results 
provide strong evidence of \ConcurGuard ’s effectiveness, 
demonstrating a greater number of successful repairs 
compared to state-of-the-art tools. 
\end{itemize}

\section{PRELIMINARIES}

In this section, we will introduce the key technologies and concepts 
employed in this paper, including the definition of the symbols 
that will be used throughout.
\subsection{Happens-Before Analysis} 

The Static Happens-Before Graph (SHBG), derived from existing work~\cite{HB, SHB}, serves as a static representation that models possible ordering constraints based on the program's structure, without relying on specific executions. This graph is essential for analyzing concurrent programs, as it captures potential ordering across different threads and aids in reasoning about synchronization and concurrency issues through static analysis of the source code. The specific definition is as follows:

\begin{definition}
\label{def:shbg}
\textbf{Static Happens-Before Graph.} The Static Happens-Before Graph of a program \( P \), denoted as \( G_{shb}(P) \), is a tuple \( \langle N, E \rangle \), where: \( N \) is a set of nodes representing events within various threads, such as memory accesses or synchronization operations. \( E \) is a set of directed edges representing the happens-before relationships inferred from the program's source code. 
\end{definition}



Unlike other approaches~\cite{Sword,race-linux,race-reverse} that incorporate synchronization-related events (e.g., locks and unlocks), our focus is limited to read and write events—and this distinction stems from the fundamental difference between our goal and that of those methods. 
Specifically, approaches like~\cite{Sword,race-linux,race-reverse} are designed for data race detection, where locks and unlocks must be carefully considered to ensure soundness (i.e., no false positives\footnote{\lnote{A \textit{false positive} refers to a tool reporting a bug that does not actually exist, while a \textit{false negative} refers to a tool failing to report a real bug.}}), 
since events that are protected by the same lock 
cannot have data races with one another.
By contrast, our task is not data race detection, but filtering out code segments irrelevant to concurrency bugs. 
For this filtering objective, lock‑protected events cannot be deemed irrelevant to concurrency bugs because misconfigurations of locks, such as inappropriate scope or granularity, and flaws in lock‑related logic often directly cause
concurrency bugs.
To maintain conservative filtering (i.e., avoiding accidental removal of code segments that may relate to concurrency bugs), we do not model locks—meaning we never filter out code segments based on lock relationships.

Given this design rationale, we formally define \textit{Event} and \textit{Conflicting Events} as follows:



\begin{definition}
\label{def:event}
\textbf{Event and Conflicting Events (CE).}  
An event $e$ is represented as a tuple $(t, s, op(m))$, where $t$ denotes
the thread ID, $s$ is a bytecode instruction 
derived 
from a statement in the program, and $op(m)$ specifies an 
operation on the memory location $m$. The operation $op$ can be  
$R$ for a read operation or $W$ for a write operation. 
Two memory access events, $a = (t_1, s_1, op_1(m))$ and 
$b = (t_2, s_2, op_2(m))$, are considered \textit{conflicting} 
(denoted $a \asymp_{CE} b$) if $t_1\neq t_2\wedge (op_1=W\vee op_2=W)$. 
\end{definition}



In the SHBG, a directed edge from one event to another indicates that the source event happens-before the target event. 
This happens-before relation is transitive: if an event happens-before a second event and the second happens-before a third, then the first happens-before the third. 
Thus, if there exists a directed path (a sequence of such edges) from $e_1$ to $e_2$, then $e_1$ happens-before $e_2$. 
Accordingly, the happens-before relation between two distinct events is determined by the presence or absence of such paths and falls into exactly one of the following three mutually exclusive and exhaustive cases:

\begin{itemize}
    \item $e_1 <_{hb} e_2$: if there exists a directed path from $e_1$ to $e_2$, then $e_1$ happens-before $e_2$.
    \item $e_2 <_{hb} e_1$: if there exists a directed path from $e_2$ to $e_1$, then $e_2$ happens-before $e_1$.
    \item $e_1 \parallel_{hb} e_2$: if neither path exists, then $e_1$ and $e_2$ are concurrent and neither happens-before the other.
\end{itemize}


In the first two cases, where a happens-before relationship exists between $e_1$ and $e_2$, 
we can also express this as $e_1\nparallel_{hb}e_2$, meaning that $e_1$ and $e_2$ are not concurrent.





\subsection{\lnote{Types and Characteristics of Concurrency Bugs}}
\label{sec:bug_types}
\subsubsection{Non-Deadlock Concurrency Bugs}%

\begin{table}[t]
 \caption{The types of memory access patterns that could potentially lead to non-deadlock concurrency bugs. 
 }
\begin{center}
\begin{tabular}{|c|p{8cm}|}
\hline
\textbf{ID} & \textbf{Memory-Access Pattern} \\
\hline
1 & $(t_a, s_i, R(x))$, $(t_b, s_j, W(x))$ \\
\hline
2 & $(t_a, s_i, W(x))$, $(t_b, s_j, R(x))$ \\
\hline
3 & $(t_a, s_i, W(x))$, $(t_b, s_j, W(x))$ \\
\hline
4 & $(t_a, s_i, R(x))$, $(t_b, s_j, W(x))$, $(t_a, s_k, R(x))$ \\
\hline
5 & $(t_a, s_i, W(x))$, $(t_b, s_j, W(x))$, $(t_a, s_k, R(x))$ \\
\hline
6 & $(t_a, s_i, W(x))$, $(t_b, s_j, R(x))$, $(t_a, s_k, W(x))$ \\
\hline
7 & $(t_a, s_i, R(x))$, $(t_b, s_j, W(x))$, $(t_a, s_k, W(x))$ \\
\hline
8 & $(t_a, s_i, W(x))$, $(t_b, s_j, W(x))$, $(t_a, s_k, W(x))$ \\
\hline
9 & $(t_a, s_i, W(x))$, $(t_b, s_j, W(x))$, $(t_b, s_k, W(y))$, $(t_a, s_l, W(y))$ \\
\hline
10 & $(t_a, s_i, W(x))$, $(t_b, s_j, W(y))$, $(t_b, s_k, W(x))$, $(t_a, s_l, W(y))$ \\
\hline
11 & $(t_a, s_i, W(x))$, $(t_b, s_j, W(y))$, $(t_a, s_k, W(y))$, $(t_b, s_l, W(x))$ \\
\hline
12 & $(t_a, s_i, W(x))$, $(t_b, s_j, R(x))$, $(t_b, s_k, R(y))$, $(t_a, s_l, W(y))$ \\
\hline
13 & $(t_a, s_i, W(x))$, $(t_b, s_j, R(y))$, $(t_b, s_k, R(x))$, $(t_a, s_l, W(y))$ \\
\hline
14 & $(t_a, s_i, R(x))$, $(t_b, s_j, W(x))$, $(t_b, s_k, W(y))$, $(t_a, s_l, R(y))$ \\
\hline
15 & $(t_a, s_i, R(x))$, $(t_b, s_j, W(y))$, $(t_b, s_k, W(x))$, $(t_a, s_l, R(y))$ \\
\hline
16 & $(t_a, s_i, R(x))$, $(t_b, s_j, W(y))$, $(t_a, s_k, R(y))$, $(t_b, s_l, W(x))$ \\
\hline
17 & $(t_a, s_i, W(x))$, $(t_b, s_j, R(y))$, $(t_a, s_k, W(y))$, $(t_b, s_l, R(x))$ \\
\hline
\end{tabular}
\end{center}
 \label{tab-pattern}
\end{table}
We define a \textit{bug-relevant memory access pattern} (hereinafter referred to as \textit{pattern}) as a specific sequence of memory access interleavings that can potentially lead to non-deadlock concurrency bugs~\cite{pattern}. Consistent with \PFIX, we adopt the 17 distinct patterns cataloged in Table~\ref{tab-pattern}, which have been proven complete~\cite{pattern}. Formally, we denote the pattern set as $\mathcal{P} = \{p_1, p_2, \dots, p_{17}\}$, where each $p_i$ corresponds to the pattern with ID $i$ in the table.

The patterns' theoretical foundation comprises two essential properties. The completeness property, derived from sequential consistency theory, guarantees that any concurrency bug arising from interleaved memory execution must conform to one of these patterns~\cite{pattern}. Formally, if a program $P$ exhibits a non-deadlock concurrency bug, then there exists a minimal event sequence $\sigma_{min} = \langle e_1, ..., e_k \rangle$ (where $k \geq 2$) in $P$ whose memory interleaving matches some pattern $p_i \in \mathcal{P}$
and which represents the
cause triggering the concurrency bug.
It should be noted, however, that the converse does not hold—the presence of a pattern indicates potential risk but does not guarantee an actual bug manifestation, since concurrent programs naturally contain numerous benign interleavings.

Furthermore, although certain patterns may appear as syntactic permutations of event orderings (e.g., patterns $p_1$ and $p_2$ in $\mathcal{P}$), they capture semantically distinct concurrency scenarios. For instance, pattern $p_1$ involves a \textit{read-write} interleaving between threads, while pattern $p_2$ involves a \textit{write-read} interleaving. These represent fundamentally different race conditions: the former may lead to lost updates, while the latter can cause stale data reads. Each scenario exhibits unique behavioral characteristics and requires specific repair strategies~\cite{pfix}.

\lnote{Among these 17 patterns, the first three ($p_1$, $p_2$, and $p_3$) result from interleavings involving exactly two memory accesses. 
These patterns are classified as either data races or order violations~\cite{bug-survey}, which differ in their underlying causes.
A data race occurs when two memory accesses may execute concurrently without being protected by the same lock. The lack of proper synchronization permits unintended interleavings, potentially leading to lost updates or stale reads. However, not all concurrency bugs caused by the interleaving of two accesses are data races~\cite{bug-survey,pattern}. Even when both accesses are protected by the same lock, and thus no data race exists, incorrect behavior may still arise if their execution order deviates from the developer’s intended ordering. Such bugs are known as order violations. An order violation occurs when two accesses are expected to execute in a specific relative order under mutual exclusion, but the actual execution violates this intended order.}

\lnote{In contrast, the remaining fourteen patterns ($p_4$ through $p_{17}$) arise from interleavings of three or more memory accesses and are representative of different forms of atomicity violations. These patterns reflect situations where a sequence of operations that should appear indivisible is improperly interleaved with accesses from other threads, potentially breaking invariants or causing inconsistent state transitions.}

\subsubsection{Deadlock Concurrency Bugs}
\lnote{Deadlock concurrency bugs are categorized into two mutually exclusive types based on competing resource properties, with distinct formation mechanisms and repair requirements~\cite{deadlock1}:}
\lnote{\begin{itemize}
\item \textbf{Resource deadlock}: Caused by cyclic competition for tangible resources (e.g., mutex locks). 
It occurs when a set of threads form a circular wait, where each thread holds at least one lock while waiting to acquire a lock held by another thread in the cycle.
\item \textbf{Communication deadlock}:
Arises from threads’ permanent blocking due to waiting for undelivered intangible signals (e.g., wait()/notify() on condition variables). Unlike resource deadlocks, this type of deadlock is unrelated to lock acquisition order.
\end{itemize}}

\lnote{Both types are semantically orthogonal to the 17 memory access patterns in \(\mathcal{P}\), as they stem from synchronization order violations rather than memory operation interleavings.}

\subsection{Large Language Models}

\begin{figure}[t!]
\centering\includegraphics[scale=0.62, trim=10cm 7.5cm 12.5cm 2.2cm, clip]{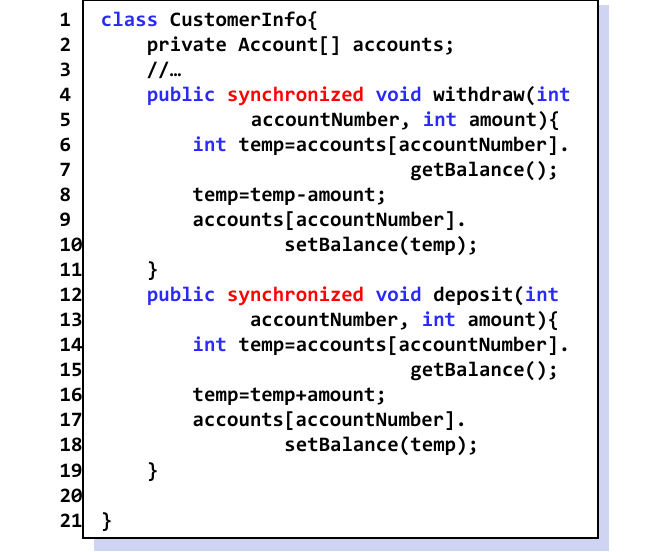}
\caption{
A Java code snippet demonstrating an atomicity violation case from the \PFIX\ benchmark~\cite{pfix}.
Concurrent executions of \texttt{withdraw()} and \texttt{deposit()} operations on the shared variable \textit{accounts} may corrupt balance fields due to unsynchronized access. The highlighted patch (red) resolves this by introducing synchronization to enforce atomicity.}
\label{fig:introduction}
\end{figure}

\subsubsection{Prompt Engineering}

LLMs, such as those in the GPT series, can have hundreds of 
billions of parameters~\cite{claude, gpt-3-5-turbo}. 
When processing a prompt, these models navigate a highly complex 
representation space, initiating auto-regressive generation 
based on the prompt's input. This richness in prompt representation 
allows for a wide variety of prompt designs, 
leading to the emergence of \textit{prompt engineering} as a new and 
evolving field~\cite{prompt}. Prompt engineering focuses on crafting 
effective prompts to guide LLMs in generating desired outputs, 
thus maximizing their performance for specific tasks.

\subsubsection{LLMs Context}
\label{llm-limitions}
Existing LLMs are initially trained on relatively short texts~\cite{gpt-3-5-turbo,Code-llama,claude}. To handle longer contexts, models such as DeepSeek-V3~\cite{deepseek-v3}, the Llama3 series~\cite{Code-llama}, and QWen2.5-Coder~\cite{Qwen} typically employ interpolation techniques like YaRN to extend their context windows from 4K/8K tokens up to 128K tokens or more~\cite{yarn}. However, merely expanding the theoretical context window does not guarantee effective practical performance on long-text tasks, as fundamental architectural limitations persist~\cite{rag,RepoAudit}.

A key challenge lies in the intrinsic limitations of the Transformer's attention mechanism when processing long sequences~\cite{transformer}. The computational complexity of standard attention grows quadratically (O(L²)) with sequence length (L), which can lead to performance degradation—as attention becomes dispersed over vast information, making it difficult to maintain focus on critical details. This phenomenon, where the model is easily distracted by irrelevant or similar information within the long context, directly undermines its ability to perform code reasoning and repair~\cite{RepoAudit,substract}.

This challenge can be intuitively understood through a human-analogous perspective—particularly since LLMs are designed to emulate human-like reasoning patterns. As shown in Figure~\ref{introduction}, when presented with well-isolated code segments containing specific bugs (e.g., the atomicity violation between \texttt{withdraw()} and \texttt{deposit()} methods), the repair task may simplify to pattern recognition, resembling how developers analyze focused code snippets. However, when the same bug is obscured within extensive irrelevant code, both humans and LLMs may face cognitive overload: essential signals can become diluted amidst irrelevant information, requiring substantial effort to distinguish causal relationships from coincidental patterns.

Therefore, to directly address the performance degradation caused by information overload in long-code contexts, our work introduces a dedicated context extractor. This approach aligns with the established practice in LLM-enabled software engineering, where extracting and condensing relevant context from the entire codebase is a fundamental step for effective reasoning~\cite{RepoAudit,agent1,agent2,agent3}. The extractor is designed to mitigate attention-related pitfalls by identifying and extracting the most bug-relevant code segments, filtering out the vast majority of irrelevant information. This enables the LLM to focus its reasoning capacity on the essential causal chain, thereby achieving more reliable APR.

\subsubsection{LLM-Driven Agents}


\begin{figure*}[t!]
\centering\includegraphics[scale=0.55, trim=2cm 5.1cm 3.5cm 1.8cm, clip]{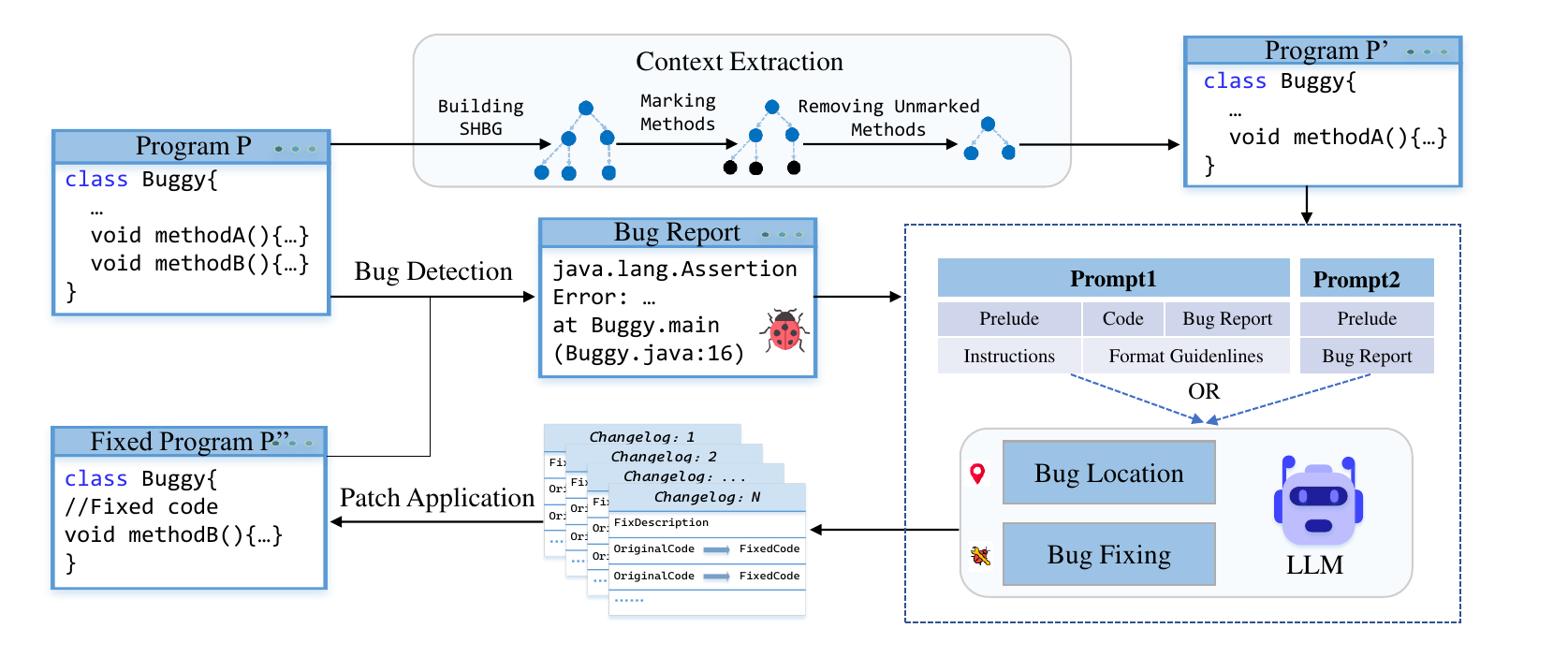}
\caption{Overview of \ConcurGuard: The input is a program containing bugs, and the output is the repaired program. The iteration stops when no concurrency bugs can be detected or when the maximum number of iterations is reached.}
\label{fig:overview}
\end{figure*}

AI agents are artificial entities capable of autonomously observing and acting upon their environments to achieve specific goals~\cite{agent3}. The concept has evolved from early implementations based on symbolic logic or reinforcement learning~\cite{agent1,agent2,agent3}, to a new paradigm driven by recent advancements in LLMs. Instead of relying solely on interactive prompts, LLM-driven agents enhance the capabilities of LLMs by enabling them to proactively perceive, reason about, and act within their environments. These agents can autonomously coordinate tasks, utilize external tools, and adapt to dynamic contexts—making them especially suitable for complex software engineering scenarios. Leveraging these capabilities, our tool is able to perform end-to-end automatic repair of concurrency bugs.

\section{\ConcurGuard}

\subsection{Overview}

Figure~\ref{fig:overview} illustrates the overall workflow of the \ConcurGuard\ framework, which starts with the bug detector identifying concurrency bugs. 
Specifically, we employ JPF~\cite{JPF} as the bug detector, which detects concurrency bugs by executing program $P$ under different thread interleavings, covering both deadlock and non-deadlock concurrency bugs. 
For non-deadlock bugs, following PFIX~\cite{pfix}, we treat any assertion violation or uncaught runtime exception during execution as a failure oracle for a bug~\cite{oracle,oracle2}, i.e., an indicator that a bug has occurred. 
This serves as the default criterion for dynamic analysis methods~\cite{ConcBugAssist,bug-widespread,failture-execution,dyn-location,mcr,JPF}, since it reflects developers' common practice of using assertions and exceptions as automated correctness checks~\cite{assert,chess}. 
Moreover, it is consistent with widely adopted concurrency bug repair datasets that rely on such observable signals as ground truth~\cite{Pecan,JaConTeBe,SIR}.

Once a concurrency bug is detected, an SHB-based context extractor then filters $P$ to retain only the most relevant code snippets $P'$, guiding the LLM's attention toward bug-related code and away from unrelated content. 
The bug reports, together with the extracted relevant code snippets, are passed to the LLM for bug localization and fixing using \texttt{Prompt1}. 
The generated patch is applied and checked by JPF. 
If the patch passes JPF checking, the process terminates successfully. 
Otherwise, \ConcurGuard\ repeats the cycle with \texttt{Prompt2}, which refines the previous output based on the JPF check result, continuing until a correct patch is produced or the maximum number of iterations is reached.

The design of the context extractor is detailed in Section~\ref{extraction}, and the bug localization and repair processes using an LLM are discussed in Section~\ref{llm}.

\subsection{Context Extraction}
\label{extraction}

\subsubsection{Problem Definition}

Before passing the source code to the LLM for bug localization and repair, \ConcurGuard\ employs a context extractor to identify and select relevant code snippets. 
We formally define the problem of selecting such pertinent snippets for LLM-based concurrency bug repair as follows:

\begin{definition}
\label{thm:extractor}
\textbf{Problem Statement.}
Given a program $P$
that contains concurrency bugs,
our context extractor $\mathcal{E}$ transforms $P$ into $P'$—where the number of tokens in $P'$
is smaller than that in $P$—while preserving critical context related to the concurrency bugs.
This reduction aims to provide more precise context to the LLM, mitigating its distraction from irrelevant code—ultimately ensuring that $CR (M(P)) \leq CR (M(P'))$, 
where $M$ denotes the output of the LLM when given the source code snippet, 
and $CR$ (repair success rate) is computed as the ratio of successful repairs to 
the total number of fixes produced. 
\end{definition}

Specifically, as part of its preprocessing steps, our context extractor removes comments from the input program $P$ by default. Notably, our evaluation results (Section~\ref{exep-extractor}) further validate the value of this design choice, demonstrating that comment filtering has a positive impact on improving repair accuracy.

\subsubsection{Key Definitions}

Before detailing our extractor approach, we first introduce two key principles that define \textit{bug-relevant event}. For each pattern $p \in \mathcal{P}$ (as listed in Table~\ref{tab-pattern}), consider a minimal event sequence $\sigma_{\text{min}} = \langle e_1, \dots, e_k \rangle$ (where k $\geq 2$ ) whose memory interleaving matches pattern $p$ and 
which represents the cause triggering the concurrency bug.
We observe that every event  $e_i \in \sigma_{\text{min}}$ has at least one corresponding conflicting event $e_j \in \sigma_{\text{min}} \setminus \{e_i\}$ satisfying $e_i \asymp_{CE} e_j$. 
This demonstrates that the presence of pairwise conflicting event pairs is necessary for the concurrency bug's manifestation.
This leads to our first principle: \textit{each memory access that potentially contributes to a concurrency bug must have at least one corresponding conflicting event in the execution trace.}

For example, consider pattern $p_4$ (ID 4 in Table~\ref{tab-pattern}) with $\sigma_{\text{min}} = \{e_1, e_2, e_3\}$ as the cause of a concurrency bug, where the sequence conforms to  $p_4$'s memory interleaving pattern. According to $p_4$'s structural constraints, we can verify that each event in $\sigma_{\text{min}}$ has a corresponding conflicting event within the sequence: $e_1 \asymp_{CE} e_2, e_2 \asymp_{CE} e_1,$ and $e_3 \asymp_{CE} e_2.$

We further establish our second principle, through logical analysis and empirical support: 
\textit{fixed-order conflicting event pairs\footnote{Fixed-order conflicting event pairs denote conflicting event pairs whose relative execution order is fixed (i.e., determined and invariant across all possible executions).} are unlikely to cause concurrency bugs.}
To understand this, we can approach the problem from the opposite perspective: if fixed-order conflicting events were capable of triggering concurrency bugs, such bugs would occur far more readily. This stands in clear contradiction to empirical studies, which indicate that most program executions complete successfully, with error-inducing thread interleavings occurring only with low probability~\cite{failure-execution}.

Based on these principles, we formally define:

\begin{definition}
\label{def:relevant}
\textbf{Bug-Relevant Event and Method}.  
In the SHBG \( \langle N, G \rangle \) of program \( P \), an event \( e \) is classified as a bug-relevant event if and only if there exists another event \( e' \in N \) such that \( e' \asymp_{CE} e \) and \( e' \parallel_{hb} e \). For a method \( m \) in program \( P \), if it contains an event \( e \) that qualifies as a bug-relevant event or includes lock events that could lead to a deadlock, then \( m \) is classified as a bug-relevant method.
\end{definition}

This definition formalizes how \ConcurGuard\ identifies and categorizes events and methods 
that may contribute to concurrency bugs. 
It is worth noting that our definition is not complete, as our second principle (\textit{fixed-order conflicting event pairs are unlikely to cause concurrency bugs}) cannot be guaranteed to hold in all cases.

Nevertheless, our second principle is reasonable, and we have not encountered any contradictions in our experiments. Furthermore, existing empirical studies on the difficulty of uncovering concurrency bugs support the notion that this principle holds true in the vast majority of cases~\cite{failture-execution,bug-diagnose}.

Bug-relevant methods include all code fragments that could potentially contribute to concurrency bugs. However, identifying these fragments alone is not sufficient. To enable effective localization and repair, it is crucial to also provide the LLM with concurrency-relevant semantics—specifically, information about thread interactions and their dependencies.
This is because without a clear understanding of inter-thread relationships, neither the LLM nor we can accurately infer which variables are shared across different threads or whether they are accessed concurrently. 
Take Figure~\ref{fig:nested_monitor} as an example, where we assume that the annotation information (i.e., the information from lines 5 and 11) is unavailable in practice. If we only rely on the given code snippet, we are essentially unable to determine whether the \texttt{put()} and \texttt{get()} methods can be executed concurrently—we have no knowledge of which threads these methods belong to, nor of the relationships between those threads.

To address this issue, we introduce the concept of semantics-relevant methods, which capture code fragments that encode or reveal important concurrency semantics. This concept is defined as follows:

\begin{definition}
\label{thm:semantics-relevant-method}
\textbf{Semantics-relevant Method.}
For a method \( m \) in program \( P \), if \( m \) invokes one or more functions responsible for managing thread lifecycles, such as thread creation, or termination, then \( m \) is classified as a semantics-relevant method. 
\end{definition}

\subsubsection{Context Extraction Based on SHBG}
\label{extractor}

\begin{figure}[t!]
\centering\includegraphics[scale=0.65, trim=10cm 3.3cm 12cm 2.2cm, clip]{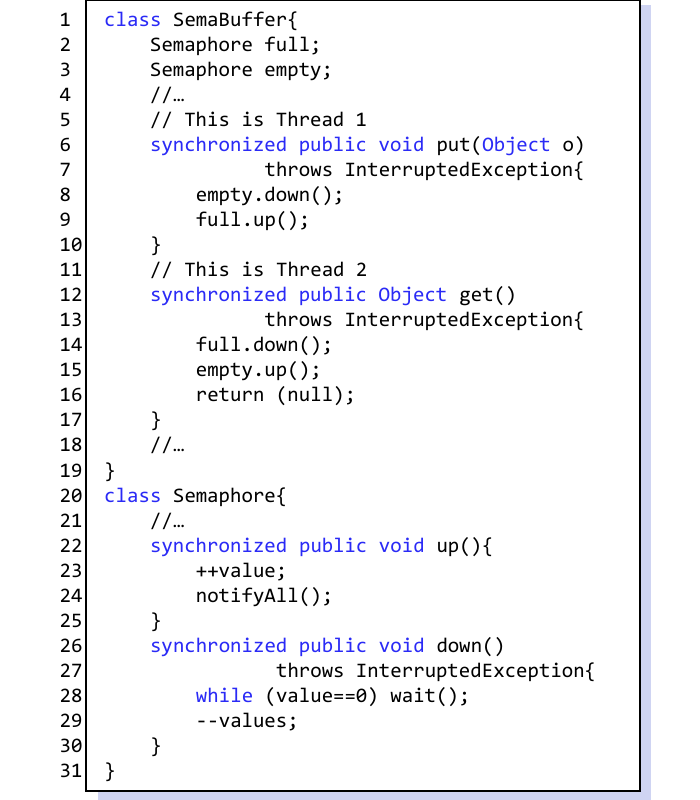}
\caption{A code snippet from the Java benchmark \texttt{nested\_monitor} in SIR~\cite{SIR}.}
\label{fig:nested_monitor}
\end{figure}

\begin{algorithm}[t!]
\caption{Identifying methods that cannot be removed.}
  \label{alg:filter}
\begin{algorithmic}[1]
    \State{\textbf{Input:} Program: $P$;}
    \State{\textbf{Output:} Marked methods set: $mSet$;}
    \State $mSet\gets \emptyset$;
    \State $\textit{SHBG}\gets \Call{ConstructSHBGraph}{P}$;

    \State $mSet\gets \Call{DectDeadlocks}{\textit{SHBG}}$;
    \State $mSet\gets mSet\cup \Call{GETSemanticMethods}{\textit{SHBG}}$; 
    \State  {\bf for }(each event $e$ in $\textit{SHBG}$)
    \State  {\hspace{1em}{\bf if} ($e$ is a \textit{bug-relevant event});}
    \State {\hspace{2em} Let $f$ be the method contain $e$;} 
    \State {\hspace{2em} $mSet \gets mSet \cup \{f\}$;}
    \State {Let $CG$ be the call graph of $P$;}
    \State {Initialize $queue$ as an empty queue;}
    \State {$checkedM\gets mSet$;}
    \State {\bf{while} ($queue \neq \emptyset$)}
    \State {\hspace{1em} $current\_node \gets \Call{dequeue}{queue}$;}
    \State {\hspace{1em} $parents\gets $ the parent nodes of $current\_node$.}

    \State {\hspace{1em}\bf{for}} (each $parent \in parents \wedge parent \notin checkedM$)
    \State {\hspace{2em} $checkedM\gets checkedM\cup parent $;}
    \State {\hspace{2em} $\Call{enqueue}{queue, parent}$;}
    \State {\textbf{return $mSet$}}
\end{algorithmic}
\end{algorithm}

As shown in Algorithm~\ref{alg:filter}, our method-marking algorithm—based on the SHBG—identifies the set of methods that cannot be removed, denoted as $mSet$. 
Any method not in $mSet$ will be filtered out later, making them invisible to the LLM.

Algorithm~\ref{alg:filter} begins by constructing the SHBG in Line~4, which comprehensively represents the happens-before relationships in program \( P \) based on its source code.
Subsequently, in Line~5, the function $\Call{DectDeadlocks}{\textit{SHBG}}$ is called to identify potential deadlocks in the program, utilizing advanced techniques from recent research~\cite{D4,Sword}. 
Line 6 invokes $\Call{GETSemanticMethods}{\textit{SHBG}}$
to identify semantics-relevant methods, which are crucial for 
understanding concurrency-related interactions within the program. 
Lines 7 through 10 identify bug-relevant methods as presented in Definition~\ref{def:relevant}. 
These methods are identified based on their potential to contribute to concurrency bugs, 
using criteria derived from the SHBG and deadlock detection results.

Finally, Algorithm~\ref{alg:filter} performs a closure computation  
in Line~11 through 19 to identify all methods that either directly 
or transitively call the identified methods in the set $mSet$.
This closure ensures that all methods potentially 
affecting the LLM's understanding and fixing of concurrency bugs are included, 
providing a comprehensive set of methods for further analysis and potential fixing.
We next provide an example to illustrate the necessity of the closure computation performed by Algorithm~\ref{alg:filter} in Lines 11 through 19.

Consider the code snippet in Figure~\ref{fig:nested_monitor}. 
The \texttt{Nested\_monitor} is a producer-consumer program from SIR~\cite{SIR} 
that contains a deadlock bug. In this program, the \texttt{put} 
method and the \texttt{get} method of the \texttt{buf} object are 
called by the producer and consumer, respectively. 
When the consumer operates faster than the producer, 
it invokes the \texttt{wait} method to block the thread. 
At this point, the lock on the \texttt{buf} object is held by the consumer, 
which in turn blocks the producer, resulting in a deadlock.

After Line 10 of Algorithm~\ref{alg:filter}, the set $mSet$ will 
include the \texttt{up()} and \texttt{down()} methods. 
These methods involve interleavings that can lead to concurrency bugs. 
However, the actual locations that require 
fixing are the \texttt{get()} and \texttt{put()} 
methods which invoke \texttt{up()} and \texttt{down()}. 
For example, removing the \texttt{synchronized} 
keyword from these methods might resolve the issue.  

If the LLM is not provided with the specific contents of 
the \texttt{get()} and \texttt{put()} methods, it may struggle to 
understand and fix these bugs effectively. To mitigate this, the closure 
computation in Lines~11 through 20 identifies and marks methods that 
transitively call the methods in $mSet$. 
This approach achieves two main objectives. 
First, it assists LLMs in pinpointing the most relevant locations for bug fixes. 
Second, it supplies additional contextual information through the call chain, 
aiding LLMs in better understanding and localizing concurrency bugs.

The time complexity for detecting the bug-related method in lines 7-10 of 
our algorithm is \(O(N^2)\). Since our detection results are at the method level, once a method is marked as bug-related, event-level analysis within that method can be terminated early, leading to faster execution times. In lines 14-19 of our algorithm, we perform a breadth-first traversal with a time complexity of \(O(N)\), where \(N\) is the number of methods in the call graph. Given that the operations at the method level are lightweight, the overall time complexity of our approach remains \(O(N^2)\). Therefore, the scalability of our Context Extractor does not limit our tool's effectiveness.

\subsubsection{Soundness and Completeness of the Context Extractor}
\label{soundAndComplete}

\lnote{We define soundness as the property that every method retained in the source code after context extraction is necessary for correctly repairing the concurrency bug. That is, the extracted context should contain only the methods essential to a correct fix.}

\lnote{Clearly, our algorithm cannot guarantee soundness. First, our \textit{bug-relevant methods} identify methods that are merely potential contributors to the interleavings leading to the concurrency bug, rather than definitively involved. Second, our method-level closure computation (Algorithm~\ref{alg:filter}, lines 14-19) is conservative: the algorithm may mark methods even if they do not contain any locations that actually require repair.}

\lnote{For completeness, we define it as the requirement that no removal of code hinders the possibility of achieving a correct repair. However, true completeness remains beyond the reach of our current approach. The sole factor limiting completeness is our second principle—namely, that \textit{fixed-order conflicting event pairs are unlikely to cause concurrency bugs}. 
It is important to note that SHB itself is complete, ensuring that all potential concurrent event relationships are captured. Additionally, the deadlock detection algorithm based on SHB does not miss any potential deadlocks~\cite{HB}.
Since the second principle cannot be guaranteed to hold in all cases, it may still lead to the omission of essential information for a correct repair. 
Nevertheless, our subsequent empirical study (Section~\ref{exep-extractor}) shows that, despite this theoretical limitation, completeness is effectively retained in practice: no benchmark repair fails solely due to omitted code snippets.}

\subsection{Bug Localization and Fixing}
\label{llm}

\subsubsection{Strategies for Utilizing LLMs in Bug Localization and Repair}

When using LLMs for bug localization and repair, we consider two distinct approaches. The first approach, denoted as $G_{\text{two\_step}}$, follows a stepwise methodology analogous to Chain-of-Thought (CoT) reasoning (refers to a reasoning strategy that decomposes a complex task into intermediate steps, allowing a model to explicitly generate intermediate reasoning traces before arriving at a final answer)~\cite{cot}. It first localizes and explains the bug, then generates a repair in separate steps, thereby enhancing reasoning transparency and facilitating validation. The second approach, denoted as $G_{\text{one\_step}}$, treats localization and repair as an atomic operation where the LLM directly generates the final patch, thereby streamlining the workflow.

After experimental evaluation (see Section~\ref{oneortwo}), \ConcurGuard\ adopts the  $G_{\text{one\_step}}$ approach. This method reduces the number of required API calls to the LLM, improving overall efficiency. More importantly, it mitigates the risk of distracting the LLM with excessive contextual information—such as detailed localization results, code snippets, and bug reports—which can hinder its focus on the core repair task~\cite{substract}.

Notably, our work is not the first to observe that using the CoT strategy is inefficient in such scenarios~\cite{evaluate}. Prior studies have also evaluated and highlighted the limitations of the CoT strategy for long-text inputs. By combining localization and repair into a single operation, \ConcurGuard\ enables the LLM to concentrate on generating appropriate fixes, thereby enhancing the efficiency and precision of the repair process.

\begin{figure}
    \centering
    \includegraphics[scale=0.47, trim=1.05cm 5.8cm 1cm 3.5cm, clip]{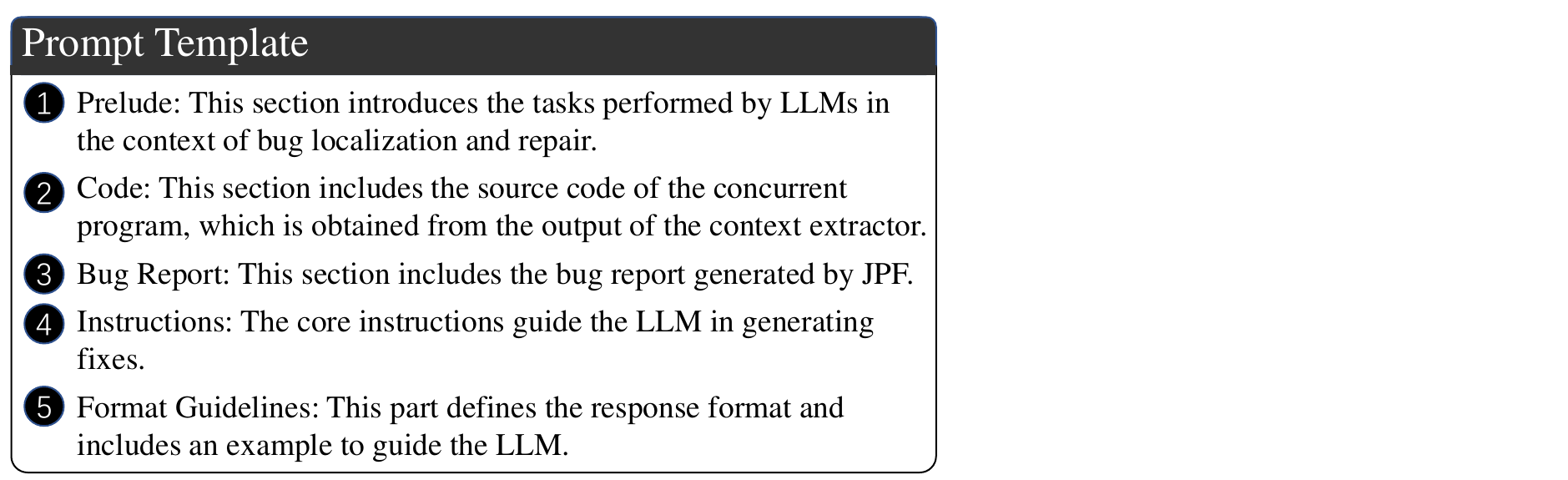}
    \caption{Prompt for Bug Localization and Fixing.}    
    \label{fig:prompt}
\end{figure}

\subsubsection{Prompt for Bug Localization and Fixing}

\begin{figure}
    \centering
    \includegraphics[scale=0.56, trim=9.1cm 7cm 1cm 4.3cm, clip]{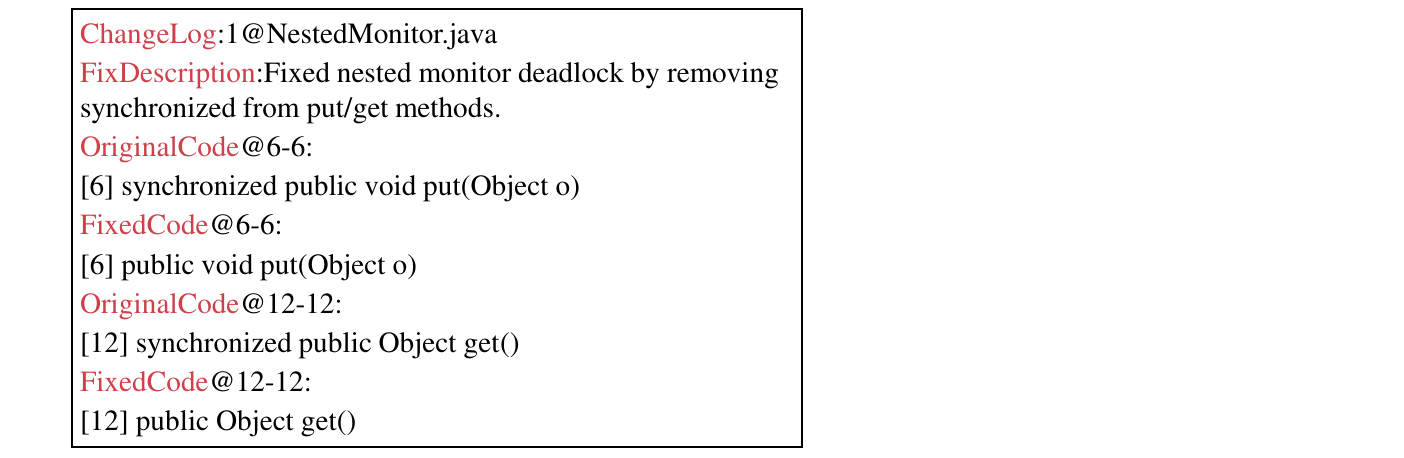}
    \caption{The output results of the LLM for the example in Figure~\ref{fig:nested_monitor}.}    
    \label{fig:example}
\end{figure}

For each input concurrent program, \ConcurGuard\ generates a structured prompt for the LLM, which consists of five components, as illustrated in Figure~\ref{fig:prompt} (see Appendix for the specific content of each component).
Among these components, the \texttt{Prelude} introduces the task as follows: 
\textit{You are given a code snippet that may contain concurrency bugs, along with a bug report detected by JPF. Some methods unrelated to the bugs have already been filtered out.}
The core \texttt{Instructions} are:  
\textit{Please fix the concurrency bug.  
Please note that some methods have been omitted. These methods are unrelated to concurrency bugs; performing any modifications to them would be considered violations. Please ensure that fixing doesn't introduce new bugs, such as deadlocks. Do not attempt to change the functionality of any function, and do not modify any code that is unrelated to concurrency bugs.}
The last part of the prompt clarifies that the format of the LLM's response
should adhere to the standards outlined in reference~\cite{response-format}, 
facilitating the parsing and application of patches to the source code (see example output in Figure~\ref{fig:example}). 
These instructions are carefully crafted to ensure that the LLM focuses exclusively on resolving the concurrency issue without introducing unintended changes or additional bugs.

The default prompt adopted in our approach, as mentioned above, was designed based on our empirical research. We also attempted to use a more concise prompt by replacing the \texttt{Prelude} and \texttt{Instructions} with a direct command: \textit{Does this concurrent program have any concurrency bugs? If yes, please fix them.} However, this modified prompt, referred to as $G_{\text{direct}}$, resulted in a significant decrease in repair accuracy (see the ablation study in Section~\ref{prompt-content} for details).

\subsubsection{Refine Outputs Based on Feedback}

After the LLM generates a fix, \ConcurGuard\ applies it to the original program. However, this fix may sometimes be invalid, for example, due to syntax errors, such as mismatched parentheses, which could cause the fix to fail compiler checks. Additionally, even if the fix passes compiler checks, it may still be incorrect and unable to pass JPF's detection.

Building on research that shows LLMs can improve their outputs or adapt based on feedback from external tools or human input—demonstrating the significant impact of these tools on enhancing model performance~\cite{llm-refine, self-refine}—we incorporate compiler or JPF error messages to refine the generated fixes.

If a generated patch fails, we prompt the LLM with the following message:  
\textit{Your patch introduced an error.  
Please review the report  below and revise your fix accordingly.  
The error report is as follows: $<$error information$>$}.

In our experimental setup, we allow a maximum of five repair attempts by the LLM. If the fix 
fails to pass the compiler or JPF’s checks after these attempts, it is considered a repair failure. 
This iterative refinement process ensures that the LLM's patches are syntactically correct and reduces 
the likelihood of introducing further errors.
\section{Evaluation}
\label{sec:ev}

We have implemented \ConcurGuard, with the static analysis component developed in Java using the Sword framework~\cite{Sword}. Sword utilizes WALA~\cite{wala} to transform the program into SSA-form IR, perform points-to analysis and construct SHBG, and is capable of deadlock detection based on SHBG.
The interaction with LLMs, as well as the processing of source code, is handled through Python code. During the evaluation, all final patches generated by \ConcurGuard\ are manually reviewed to ensure they do not unintentionally alter the intended functionality of the program. This prevents incorrect fixes from being mistakenly counted as successful repairs. Additionally, in the bug detection phase, we adopt the same setup as \PFIX~\cite{pfix} for evaluating patch correctness—specifically, running the repaired program 100 times under random thread scheduling. In the following sections, we will elaborate on the evaluation setup and present our experimental results. Our experiments were conducted on a 12th Gen Intel(R) Core(TM) i7-12700 with 32GB of memory.
\subsection{Evaluation Settings}

\subsubsection{Research Questions}
We aim to answer these research questions:
\begin{itemize}
\item RQ1: How effective is \ConcurGuard\ in correctly fixing concurrency bugs?
\item RQ2: What is the capability of the context extractor in filtering irrelevant code, 
         and its impact on repair results?
\item RQ3: What is the impact of prompting strategies on the performance of concurrency bug repair?
\item RQ4: What is the quality of the fixes provided by \ConcurGuard?
\end{itemize}

\subsubsection{LLMs Configuration}

We apply $top\_p=1$ to enforce deterministic decoding by consistently selecting the most probable token at each generation step, following~\cite{response-format,quixbugs-repair}. A low temperature of 0.2—empirically shown to maximize pass@1\footnote{The pass@k metric measures the probability that at least one of k samples passes all tests; pass@1 refers to the proportion of problems solved correctly on the first generation attempt.} performance~\cite{code-completion}—is adopted following~\cite{LLM-FIX2,code-completion,tempeture,response-format}. 
All other parameters, including $frequency\_penalty$ and $presence\_penalty$, are set to the model's default values, which aligns with common practice~\cite{LLM-FIX2,code-completion,tempeture,response-format,quixbugs-repair,LLM-FIX1}.

We employed the GPT-4 series, specifically gpt-4-turbo-2024-04-09~\cite{gpt-4-turbo}, since it offers strong reasoning capabilities. In the ablation study, we also evaluated the GPT-3.5 series, specifically gpt-3.5-turbo-0125~\cite{gpt-3-5-turbo}. Both models produced consistent conclusions. For brevity, unless otherwise noted, subsequent discussions will primarily reference results from GPT-4.

\subsubsection{Baselines}

\PFIX\ and \HIPPODROME\ represent 
the state-of-the-art in APR for concurrency bugs, 
utilizing dynamic and static analysis methods, respectively.
However, \ConcurGuard\ includes deadlock repair capabilities, 
which are not covered by either \PFIX\ or \HIPPODROME~\cite{hipp}. 
Despite our efforts, we were unable to find any open-source tools specifically designed for deadlock repair~\cite{deadlock1,deadlock2,dfix}.

We also considered comparing \ConcurGuard\ with InferFix~\cite{Inferfix}, an end-to-end LLM-based repair tool capable of addressing TSVs. Unfortunately, its source code is not publicly available. Fortunately,  the set of concurrency bugs targeted by InferFix are identical to those of \HIPPODROME, 
and \HIPPODROME\ further provides a theoretical guarantee for the correctness of the fixes it generates~\cite{hipp}.
Given this overlap in bug scope and \HIPPODROME’s additional correctness guarantee, we believe a direct comparison with InferFix is unnecessary.

We also included recent agent-based repair techniques capable of autonomously fixing bugs with the aid of human-provided context, such as GitHub issue discussions~\cite{SWE,agentless,Autocoderover}. These techniques are designed to address a wide range of bug types, including concurrency bugs. 
Among these candidates, we selected SWE-agent~\cite{SWE} as a comparative baseline. Notably, the techniques proposed in~\cite{agentless,Autocoderover} were excluded from our baseline comparisons, primarily because they do not provide support for Java—the target programming language of \ConcurGuard. 
For SWE-agent, we supplied bug reports generated by JPF and interacted with the tool via its web-based service under default parameter settings. 
Experimental results indicate, however, that its repair effectiveness is lower than that of \PFIX\ and \HIPPODROME. Specifically, when using GPT-4, SWE-agent achieves an accuracy of 62\%, compared to 67\% for \PFIX\ and 64\% for \HIPPODROME\ (detailed results are provided in Tables~\ref{tab:agent} and~\ref{tab:compare}). Given this performance gap, our primary comparative analysis focuses on traditional program repair methods.

\subsubsection{Benchmark}
\label{benchmark}
Our benchmark selection follows the same sources as \PFIX\ and \HIPPODROME, drawing from the SIR repository~\cite{SIR}, Pecan's benchmark programs~\cite{Pecan}, and JaConTeBe~\cite{JaConTeBe}. However, our methodology requires the exclusion of benchmarks that satisfy any of the following criteria: 1) lack relevance to concurrency interleavings; 2) contain bug manifestations outside source code boundaries, as our technique operates at the source code level; 3) exceed the LLM's token processing constraints; or 4) surpass JPF's detection capabilities. These constraints led to the exclusion of several JaConTeBe benchmarks employed in \HIPPODROME—specifically, one from \textit{dbcp}, three from \textit{groovy}, and ten from \textit{jdk}. Within this excluded set, \textit{dbcp\_1}, \textit{groovy\_6}, \textit{jdk6\_1}, and \textit{jdk6\_4} were removed due to external bug locations, while the remainder were excluded because JPF could not identify the bugs. Notably, \HIPPODROME\ successfully repaired four of these excluded cases (\textit{jdk6\_1}, \textit{jdk6\_5}, \textit{jdk7\_1}, and \textit{log4j3}). However, even with their inclusion, \HIPPODROME's correct repair count would remain inferior to \ConcurGuard's. Our final benchmark suite incorporates all PFIX benchmarks and represents diverse concurrency bug categories—including atomicity violations, data races, and deadlocks—totaling 33 non-deadlock concurrency bugs and 10 deadlock instances.

We also explored existing deep learning datasets~\cite{MODIT,mt-fix,CPatMiner,Dlfix,Inferfix} for benchmarks on concurrency bugs. Unfortunately, these datasets consist solely of code snippets related to concurrency bugs and lack the necessary test suites for dynamic execution, which does not align with our end-to-end objectives.

\begin{table*}
  \centering
  \caption{Results of Comparing \ConcurGuard\ with \PFIX\ and \HIPPODROME\ on a Set of 
  Non-Deadlock Concurrency Bugs.
Here, Lock denotes the number of locks introduced in correctly repaired instances. Time in \ConcurGuard\ represents the total time for \ConcurGuard\ execution, excluding the time spent on API calls to the LLM.}
   \label{tab:compare}
   \resizebox{\textwidth}{!}{ 
  \begin{tabular}{ccccccccccccc}
    \toprule
    \multicolumn{3}{c}{\centering Subject Apps} & \multicolumn{3}{c}{\PFIX} &  \multicolumn{3}{c}{\HIPPODROME} & \multicolumn{4}{c}{\ConcurGuard} \\
    Program Name   & Loc(Token) &Bug Type   & Correct?   & Lock  &Time(s)&Correct?& Lock &Time(s)&Correct? &  Iter & Lock &Time(s)\\
    \midrule
    account        & 102(1030) & \lnote{atomicity violation} & ×          &1  &-     &×         &-    &-    &\checkmark  &1  &1 &1.54 \\
    accountsubtype & 138(1087) & \lnote{atomicity violation} & \checkmark &1  &80.24 &×         &-    &-    &\checkmark  &1  &1 &2.74 \\
    airline        & 51(609)  & \lnote{atomicity violation} & \checkmark &1  &24.55 &\checkmark &6   &2.05 &\checkmark  &1  &2 &1.19 \\
    alarmclock     & 206(2141) &\lnote{data race}& ×          &-  &-     &\checkmark &22  &6.58 &\checkmark  &3  &8 &6.28 \\
    atmoerror      & 48(353)  &\lnote{data race} & \checkmark &2  &13.25 &\checkmark &5   &1.88 &\checkmark  &1  &2 &0.88 \\
    buggyprogram   & 258(2330) &\lnote{atomicity violation} & \checkmark &3  &43    &×          &-   &-    &\checkmark  &1  &2 &6.01 \\
    checkfield     & 41(248)  &\lnote{atomicity violation} & \checkmark &2  &16.95 &×          &-   &-    &\checkmark  &1  &2 &5.4  \\
    consisitency   & 28(203)  &\lnote{atomicity violation} & \checkmark &2&16.7&\checkmark &7 &3.17 &\checkmark  &1  &1 &0.83 \\
    critical       & 56(349)  &\lnote{atomicity violation}  &\checkmark &2  &29.5   &\checkmark &6  &6.95 &\checkmark  &2  &2 &7.11 \\ 
    datarace       & 90(625)  &\lnote{data race}       & \checkmark &2 &59.25  &\checkmark &2  &1.56 &\checkmark    &1  &3 &4.13 \\
    derby3         & 129(715) &\lnote{data race} &×  &- &- &\checkmark &6   &4.56 &\checkmark  &1  &2 &0.54 \\
    elevator       & 1192(9430)&\lnote{data race} & ×          &-  &-     &\checkmark &16 &8.56 &\checkmark  &1  &6 &10.56\\
    even           & 49(275)  &\lnote{atomicity violation}& \checkmark &1 &98.4   &\checkmark &5  &2.5  &\checkmark  &1  &1 &3.98 \\
    groovy1        & 124(955) &\lnote{data race}& ×          &-  &-     &\checkmark &10 &6.74 &\checkmark  &1   &1  &0.41 \\
    groovy3        & 87(265)  &\lnote{order  violation} & ×          &-  &-     &×          &-  &-    &\checkmark  &1  &1  &0.52 \\
    hashcodetest   & 1258(20269)&\lnote{atomicity violation} &\checkmark  &1 &15.9   &\checkmark &12 &6.27 &\checkmark  &1  &0 &4.42 \\
    log4j          &18799(220339)&\lnote{atomicity violation}  &\checkmark  &1 &43.25  &\checkmark &12 &3.17 &\checkmark  &1  &1 &3.21 \\
    log4j1         & 81(530)  &\lnote{atomicity violation} &\checkmark  &1 &42.15  &\checkmark &12 &3.17 &\checkmark  &1  &1 &3.03 \\
    linkedlist     & 204(1591) &\lnote{atomicity violation}   &\checkmark  &1 &43.2   &\checkmark &4  &2.08 &\checkmark  &1  &1 &4.82 \\
    mergesort      & 270(2712) &\lnote{data race}&×           &- &-      &\checkmark &39 &3.94 &\checkmark  &2  &6 &7.1 \\
    pingpong       & 130(1216) &\lnote{order violation}&\checkmark  &0 &48.25  &\checkmark &4 &5.15 &\checkmark  &1  &1 &10.52\\
    pool           & 1815(27137)&\lnote{atomicity violation} &×           &- &-      &×          &-  &-    &×           &-  &- &-    \\
    ProducerConsumer& 144(985)&\lnote{order violation}&×         &- &-      &×          &-  &-    &×           &-  &- &-    \\
    raxextended     & 222(968)&\lnote{atomicity violation}  &\checkmark  &2  &-      &×          &-  &-   &\checkmark  &1  &5 &10.08\\
    raxextended2    & 222(970)&\lnote{atomicity violation}  &\checkmark  &2  &-      &×          &-  &-   &×          &-   &- &-    \\
    reorder2        & 135(532)&\lnote{atomicity violation} &\checkmark &1&19.6&×           &-  &-    &\checkmark &1   &0 &4.17 \\
    store           &56(283)  &\lnote{atomicity violation}  &\checkmark  &2  &13.05 &\checkmark  &2  &1.51 &\checkmark &1   &1 &6.15 \\
    stringbuffer     &416(12467)&\lnote{atomicity violation} &\checkmark  &1  &29.2  &×           &-  &-    &×          &-   &- &-\\
    test1            &35(166) &\lnote{atomicity violation} &\checkmark  &2  &12.1&\checkmark    &2  &1.56 &\checkmark &1   &2 &4.07\\
    test2            &35(168) &\lnote{atomicity violation}  &×           &-  &-&\checkmark       &2  &1.60 &\checkmark &1   &2 &3.85\\
    twoStage         &137(775)&\lnote{atomicity violation}  &×           &-  &-&×                &-  &-    &\checkmark &2   &1 &2.4 \\
    wrongLock        &112(589)&\lnote{atomicity violation}  &\checkmark  &1  &13.05&\checkmark   &1  &1.60 &\checkmark &1   &1 &4.23\\
    wrongLock2       &50(238) &\lnote{data race}      &\checkmark   &1 &23.7 &\checkmark   &2  &2.33 &\checkmark &1   &1 &3.79\\
    \midrule
    \multicolumn{3}{c}{\centering Total correctly fixed}&22 & & &21  & & &29& & \\
    \bottomrule
  \end{tabular}}
\end{table*}

\begin{table}
  \centering
  \caption{\lnote{Results of \ConcurGuard\ in Fixing Deadlock, where \textit{d1}, \textit{d2}, and \textit{diningPhilosophers} belong to resource deadlocks, while the remaining programs belong to communication deadlocks.}}
  \label{tab:deadlock}
  \begin{tabular}{cccccc}
    \toprule
    \multicolumn{2}{c}{\centering Subject Apps} & \multicolumn{4}{c}{\ConcurGuard} \\
    Program Name   & Loc(Token)& Correct? &  Iter & Lock &Time(s)\\
    \midrule
    clean            & 113(612) &×            &-   &- &-\\
    d1               &42(183)   &\checkmark   &1   &0 &4\\
    d2               &48(285)   &\checkmark   &1   &0 &3.84\\
    diningPhiliosophers&65(443) &\checkmark   &1   &0 &4.05\\
    groovy         & 363(2356)   &×            &-   &- &-  \\
    loseNotify     & 115(559)   &×            &-   &- &-  \\
    nested\_monitor& 126(614)   &\checkmark   &4   &0 &5.31\\
    readers\_writers & 196(1248) &×            &-   &- &-\\
    replicated\_workers&672(4298)&×            &-   &- &-\\
    sleepingBarber   &155(891)  &×            &-   &- &-\\
    \bottomrule
  \end{tabular}
\end{table}
\subsection{RQ1: Effectiveness of \ConcurGuard}
\label{sec:compare}
\paragraph{Comparison with \PFIX\ and \HIPPODROME}
Table~\ref{tab:compare} and Table~\ref{tab:deadlock} present the repair results of \ConcurGuard\ on non-deadlock 
and deadlock concurrency bugs, respectively. Table~\ref{tab:compare} also 
includes a comparison of \ConcurGuard, \PFIX, and \HIPPODROME\ in fixing non-deadlock concurrency bugs. Under \texttt{Subject Apps}, basic information about the tested programs is provided, 
with \texttt{Loc(Token)} representing lines of native code and the number of tokens (excluding library calls).
Note that the source code of dependent JAR packages for these subject apps is not counted in \texttt{Loc}.
These libraries are processed during both the dynamic execution phase (JPF) and the static analysis phase (context extractor) to ensure successful compilation, but their contents are not visible to the LLM, which performs analysis exclusively at the source code level of the target project.
Specifically, the sizes of dependent JAR packages for \textit{derby3}, \textit{groovy1}, \textit{groovy3}, \textit{pool}, and \textit{groovy} are 4220 KB, 5584 KB, 5584 KB, 276 KB, and 5584 KB, respectively. 
Columns labeled \texttt{Correct?} indicate whether the repairs generated by \PFIX, \HIPPODROME,\footnote{Some repair results differ from those reported in~\cite{hipp} due to variations in the oracles. As a static analysis-based repair method, \HIPPODROME\ addresses only potential data races without considering whether those data races are the root cause of the concurrency bugs. In contrast, our oracle emphasizes whether the crash has been effectively resolved.} or \ConcurGuard\ were correct. The \texttt{Time} columns show execution times, with \ConcurGuard's execution time excluding time spent on API calls to the LLM. Finally, the \texttt{Iter} columns 
reports the number of iterative API calls to the LLM made by \ConcurGuard.
Overall, \ConcurGuard\ successfully fixes 29 non-deadlock bugs, 
outperforming \PFIX\ (22) and \HIPPODROME\ (21). Additionally, \ConcurGuard\ successfully repaired 4 deadlock concurrency bugs,  
a type of bug that neither \PFIX\ nor \HIPPODROME\ is capable of addressing.  
These results demonstrate that \ConcurGuard\ outperforms the state-of-the-art approaches in effectively repairing various types of concurrency bugs.

While \PFIX\ demonstrated some success, it faced challenges in several benchmarks. For example, it successfully identified the memory access pattern triggering the bug in \textit{account}, but the repair process inadvertently introduced a deadlock. Similarly, in \textit{alarmclock}, a deadlock occurred as part of the repair. For 9 additional benchmarks, \PFIX\ struggled to correctly identify the memory access patterns causing the bugs, which limited its effectiveness. \HIPPODROME, on the other hand, is adept at repairing data race issues but is not designed to address concurrency bugs involving complex memory interleavings~\cite{hipp}. We found that the data races repaired by \HIPPODROME\ were not the root cause of the concurrency bugs, and it also introduced excessive locking to protect variables that do not need protection.

Although \ConcurGuard\ achieved strong overall results, it did not succeed in repairing four non-deadlock concurrency bugs and six deadlock concurrency bugs. Analysis revealed that these failures were primarily due to difficulties in localizing the concurrency bugs. In some cases, the generated repairs altered the original functionality or introduced inaccuracies, such as in the \textit{producerConsumer} benchmark, where the repair incorrectly increased the number of threads. These challenges suggest opportunities for further refinement, particularly in improving the precision of bug localization. However, for \textit{loseNotify}, the LLM successfully pinpointed the source of the concurrency bug, though the resulting repair did not fully guarantee deadlock elimination.

\paragraph{Comparison with SWE-agent}

Table~\ref{tab:agent} presents the comparison results between \ConcurGuard\ and SWE-agent after five independent runs of all benchmarks, with averages calculated to minimize randomness. Columns \(A_{\text{location}}\) and \(A_{\text{fix}}\) represent the success rates of localization and repair, respectively. Results for both GPT-3.5 and GPT-4.0 are presented.

To determine whether the LLM has accurately localized a concurrency bug based on its output, our criterion is whether the LLM's repair description includes the actual cause of the concurrency bug and does not include any potentially functional-altering descriptions. We acknowledge that this assessment is subject to inherent subjectivity, which arises from two principal sources: (1) explanatory ambiguity: concurrency bugs often admit multiple valid root cause explanations~\cite{bug-diagnose}. A data race, for example, can be attributed to a missing lock or to a violation of execution order—both are technically correct but emphasize different facets of the fault. (2) interpretive judgment: assessing the LLM's diagnostic text requires matching its phrasing to a conceptual standard. For instance, a description such as \textit{uncontrolled concurrent access} may be accepted by some evaluators as a valid characterization of a race condition, yet rejected by others for lacking technical precision. 
Despite relying on votes from five evaluators to determine outcomes, variability in their judgments still persists. Therefore, we posit that \(A_{\text{fix}}\) serves as a more objective and reliable metric, 
as the correctness of the generated patch program can be unequivocally confirmed through dynamic execution.

Across five independent runs, SWE-agent achieved successful repairs for only 15 non-deadlock benchmarks—where a benchmark is considered successfully repaired if a fix is achieved in at least one run, fewer than both \PFIX\ and \HIPPODROME. The individual run results for all deadlock and non-deadlock benchmarks (17, 18, 17, 16, 18 successful repairs) yield an average success rate of merely 40\%, substantially lower than \ConcurGuard's 76\% success rate. 
Notably, the maximum number of successful repairs achieved by SWE-agent (18) is even lower than the minimum number of successful repairs (27) achieved by \ConcurGuard\ across its five independent runs. Furthermore, SWE-agent failed to successfully repair any benchmarks that \ConcurGuard\ could not resolve. These results suggest that the global and non-localized nature of concurrency bugs may not be adequately captured by existing agent-based techniques, which likely overlook key concurrency semantics and event-ordering constraints, thereby limiting their effectiveness in LLM-based repair scenarios.

\begin{table}
  \centering
  \caption{Comparison results with SWE-agent.}
  \label{tab:agent}
  \begin{tabular}{ccccc}
    \toprule
\multirow{2}{*}{Method} & \multicolumn{2}{c}{GPT3.5}& \multicolumn{2}{c}{GPT4.0} \\
       & $A_{location}$ & $A_{fix}$ &  $A_{location}$ &  $A_{fix}$\\
    \midrule
    SWE-agent          &28\% &21\%        &62\% &40\%   \\
   \ConcurGuard        &\textbf{36\%} &\textbf{36\%}   &\textbf{79\%} &\textbf{76\%}   \\
    \bottomrule
  \end{tabular}
\end{table}

\begin{resultbox}
    \textbf{Answer to RQ1}: 
    Our results show that \ConcurGuard\ outperforms state-of-the-art approaches in repairing concurrency bugs. It fixes 29 non-deadlock bugs, more than \PFIX\ (22) and \HIPPODROME\ (21), and uniquely repairs 4 deadlock bugs that neither \PFIX\ nor \HIPPODROME\ can handle. Compared to the agent-based SWE-agent using GPT-4, \ConcurGuard\ achieves a 76\% repair success rate versus SWE-agent’s 40\%. These results demonstrate \ConcurGuard’s strong capability across diverse concurrency bug types.
\end{resultbox}

\subsection{RQ2: Context Extractor's Filtering Capability and Repair Impact}
\label{exep-extractor}

Initially, \ConcurGuard\ begins with standard preprocessing by removing comments. Following this, it filters out methods that have not been marked according to Algorithm~\ref{alg:filter}. It is important to note that the SHBG is constructed based on the call graph. Therefore, methods that are not present in the 
call graph are filtered out before the construction of the SHBG in Algorithm~\ref{alg:filter}. We abstract the context extraction process into three stages: first, removing comments; second, excluding code that is outside the call graph; and third, filtering out code not included in $mSet$ based on the SHBG. The original program source code is denoted as $P_{original}$, with the outputs of these three stages represented as $P_{comment}$, $P_{call\_graph}$, and $P_{shb}$, respectively.
To simplify the notation, we also denote $P_{original}$, $P_{comment}$, $P_{call\_graph}$, and $P_{shb}$ as $P_1$, $P_2$, $P_3$, and $P_4$, respectively.
\lnote{To quantify the unsoundness of the context extractor, we manually remove the irrelevant methods described in Section~\ref{soundAndComplete} from $P_{shb}$ to obtain an ideal filtered version $P_{\text{ideal}}$ ($P_5$).}

\begin{table}
  \centering
  \caption{\lnote{Filtering Capabilities at Different Stages and Their Impact on Repair Accuracy Rate.}}
  \label{tab:filter}
  \begin{tabular}{cccccc}
    \toprule
    \multirow{2}{*}{Input} & \multicolumn{2}{c}{GPT3.5}& \multicolumn{2}{c}{GPT4.0}& $\%_{filtered}$ \\
       & $A_{location}$ & $A_{fix}$ &  $A_{location}$ &  $A_{fix}$\\
    \midrule
    $P_{original}$       &28\% &27\%        &67\% &62\%   &-\\
    $P_{comment}$        &30\% &29\%        &70\% &65\%   &49\%\\
    $P_{call\_graph}$    &32\% &32\%        &72\% &70\%   &75\%\\
    $P_{shb}$            &36\% &36\% &79\% &76\%   &18\%\\
    $P_{ideal}$          &\textbf{38\%} &\textbf{37\%} &\textbf{82\%} &\textbf{79\%}   &15\%\\
    \bottomrule
  \end{tabular}
\end{table}

\begin{figure}
    \centering
    \includegraphics[scale=0.45, trim=10cm 9.2cm 10.5cm 4.3cm, clip]{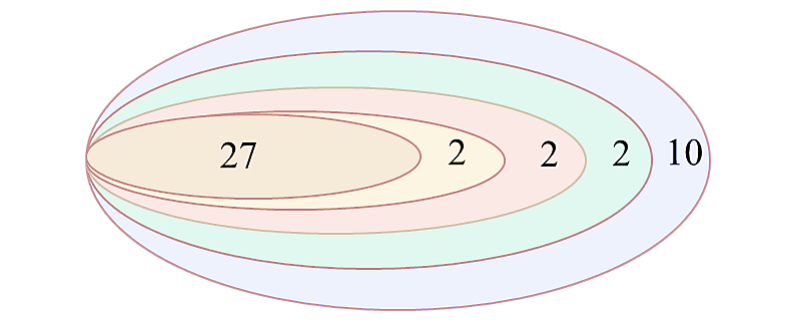}
    \caption{Number of successfully repaired (where success is defined as at least one successful repair in five independent runs) benchmarks across different filtering stages.
From innermost to outermost, the circles represent successful repair counts under the $P_{\text{original}}$,  $P_{\text{comment}}$, $P_{\text{call\_graph}}$, and  $P_{\text{shb}}$ stages, with the largest circle indicating the total benchmark set size.  
This visualization demonstrates how each filtering stage incrementally expands the set of repairable benchmarks.}
    \label{fig:intersection}
\end{figure}

In Table~\ref{tab:filter}, we present the filtering capabilities at each stage and their impact on repair results after independently executing all programs five times. The final column shows the average proportion of tokens filtered out at each stage, calculated as \(1 - \frac{P_i}{P_{i-1}}\).

From Table~\ref{tab:filter}, we observe that for both GPT-3.5 and GPT-4.0, repair accuracy increases as the number of filtered tokens rises. Notably, when using the GPT-4.0 model, applying $P_4$ yields a 14\% accuracy improvement compared to $P_1$. This finding further confirms that redundant tokens undermine the effectiveness of LLMs~\cite{substract,evaluate}.  
\lnote{Moreover, although our context extractor does not achieve soundness---meaning there remains room to further filter approximately 15\% of the tokens
in its current output---it nearly attains the accuracy that would be achieved with full soundness, exhibiting only a 3\% drop in repair accuracy.}

\lnote{To further evaluate the incompleteness of our context extractor, we conducted five independent runs for each program under each of the four sequential processing states (\( P_1 \), \( P_2 \), \( P_3 \), \( P_4 \)) to test whether our filtering process causes any concurrency bugs to fail to be repaired.}
Figure~\ref{fig:intersection} illustrates the number of benchmarks successfully repaired (a benchmark is considered successful if it achieves a successful repair in at least one of five independent runs) across our four sequential filtering stages. The nested structure demonstrates that the set of repairable benchmarks progressively expands from $P_1$ to $P_4$ (from the innermost circle outward), with the outermost circle representing the total benchmark set.

\lnote{Our experiments reveal a critical finding: throughout all five independent runs, no concurrency bugs that were repairable under $P_1$ failed to be repaired under $P_4$. Furthermore, we inspected all benchmarks and empirically confirmed that our second principle---fixed-order conflicting event pairs are unlikely to cause concurrency bugs---holds in practice. These results indicate that although our context extractor is not theoretically complete, it achieves completeness in practice on our benchmark suite. Moreover, each filtering stage builds constructively on the previous one---each stage contributes to improving repair accuracy by helping the LLM narrow its focus to the code segments most relevant to bug localization and subsequent repair.}

\begin{resultbox}
    \textbf{Answer to RQ2}: 
Our results demonstrate that the context extractor significantly enhances repair accuracy by effectively filtering out irrelevant code. The multi-stage filtering process not only reduces the amount of noise in the input data but also allows the LLM to focus on critical sections of code pertinent to bug localization and repair.
\lnote{Furthermore, we quantified the soundness of the context extractor and found that the gap between our method and theoretical soundness is relatively small. Regarding completeness, although our approach lacks a theoretical guarantee, the empirical results show that completeness is effectively retained in practice.}
\end{resultbox}

\subsection{RQ3: The Impact of Prompt Strategies}
\label{sec:strate}

\begin{table}
  \centering
  \caption{The impact of different strategies for utilizing LLMs on repair results is shown.}
  \label{tab:prompt}
  \begin{tabular}{cccccc}
    \toprule
       \multirow{2}{*}{Input} & \multirow{2}{*}{Startegy} & \multicolumn{2}{c}{GPT3.5}& \multicolumn{2}{c}{GPT4.0}\\
       & & $A_{location}$ & $A_{fix}$ &  $A_{location}$ &  $A_{fix}$\\
    \midrule
     \multirow{4}{*}{$P_{original}$} &$G_{one\_step}$ & 28\% &\textbf{27\%}       &67\% &\textbf{62\%}   \\
    &$G_{two\_step}$      &\textbf{30\%}  &25\% &\textbf{70\%}&60\%   \\
    &$G_{direct}$         &24\%  &22\% &51\%&40\%   \\
    &$G_{no\_bug\_info}$  &26\%  &25\% &66\%&61\%   \\
    \midrule
   \multirow{4}{*}{$P_{shb}$} &$G_{one\_step}$      &36\% &\textbf{36\%}        &79\%& \textbf{76\%}  \\
    &$G_{two\_step}$      &\textbf{37\%} &33\%       &\textbf{81\%}& 65\%  \\
    &$G_{direct}$         &32\% &30\%       &60\%& 58\%  \\
    &$G_{no\_bug\_info}$  &36\% &35\%       &77\%& 76\%  \\
    \bottomrule
  \end{tabular}
\end{table}

In this section, we first investigate whether separating the localization and repair processes, 
using the CoT~\cite{cot} approach, leads to better results. 
Next, we examine the impact of different prompt contents on the accuracy of the repairs.
Table~\ref{tab:prompt} illustrates the impact of different prompt strategies on repair outcomes, \lnote{showing the results for $P_{\text{original}}$ (without the context extractor) and for $P_{\text{shb}}$ (with the context extractor).}

\subsubsection{Atomic vs. Separate Localization and Repair}
\label{oneortwo}
$G_{\text{one\_step}}$ and $G_{\text{two\_step}}$ represent two distinct approaches to bug localization and repair, with $G_{\text{one\_step}}$ serving as the default approach in our current work. In $G_{\text{one\_step}}$, localization and repair are treated as a single atomic operation. In contrast, $G_{\text{two\_step}}$ separates these two tasks into distinct sequential steps.

When localization and repair are performed separately, before the LLM executes the repair, 
we first provide the LLM with a prompt to locate the bug. 
The prompt consists of four parts: (1) a description stating, \textit{You were given a program source code that may contain concurrency bugs}; (2) the program 
source code itself; (3) a bug report generated by JPF and (4) the required format for the bug report, 
which includes three components: \texttt{Bug Type} (type of the bug detected), \texttt{Bug Description} (a detailed description of the bug), and \texttt{Bug Location} 
(the location of the bug, indicated by file name and line number).

From the experimental results, we found that separating localization and repair did not improve repair accuracy, \lnote{regardless of whether the context extractor was enabled or not.} Instead, when using the GPT-4.0 model, the accuracy decreased by 11\% \lnote{with the context extractor enabled.} This decline may be attributed to the challenges posed by long contexts, which can hinder the LLM's comprehension and consequently reduce the accuracy of concurrency bug repairs~\cite{evaluate}.

\subsubsection{Impact of Prompt Content on Repair Accuracy\label{pg}}
\label{prompt-content}

The prompt $G_{\text{one\_step}}$ consists of five components, with the final part specifying the response format following the standard outlined in~\cite{response-format}. Since the impact of formatting on repair accuracy has been extensively studied in prior work~\cite{response-format}, our investigation focuses instead on the remaining components of $G_{\text{one\_step}}$. To evaluate their effectiveness, we conducted an ablation study where the original \texttt{Prelude} and \texttt{Instructions} were replaced with the simplified directive: \textit{Does this concurrent program have any concurrency bugs? If yes, please fix them.} This modified prompt, termed  $G_{\text{direct}}$, retained all other components unchanged. Experimental results (Table~\ref{tab:prompt}) show that $G_{\text{direct}}$ leads to a significant decrease in repair accuracy, confirming the effectiveness of our original prompt design in guiding the LLM's reasoning process for concurrency bug repair.

Furthermore, we conducted an ablation study by removing the bug report section from the $G_{\text{one\_step}}$ prompt, resulting in a revised version termed $G_{\text{no\_bug\_info}}$. \lnote{
This experiment evaluates the impact of JPF's incompleteness on repair outcomes, specifically the effect when JPF does not provide a detected concurrency bug report.
As shown in Table~\ref{tab:prompt}, the repair accuracy exhibits only minor changes, regardless of whether the context extractor is enabled or not.}
This result stems from the inherent redundancy of bug report information for concurrency bug repair: although such reports capture surface-level symptoms such as stack traces from triggered exceptions, they fail to reveal the root cause---such as specific thread interleavings---which is often spatially and temporally distant from the observable failure. Moreover, the key information contained in bug reports (i.e., stack traces from triggered exceptions) can typically be inferred directly from the source code through existing error-handling constructs like \texttt{throw} or \texttt{assert} statements, further diminishing the value of explicit bug reports. This outcome aligns with the recognized difficulty in diagnosing concurrency bugs even when bug-triggering traces are available, as locating the root cause remains challenging~\cite{pfix}.

\begin{resultbox}
    \textbf{Answer to RQ3}: Our experimental results indicate that treating localization and repair as an atomic operation outperforms separating the two processes. Additionally, our evaluation of different prompt contents shows that carefully crafted prompts significantly enhance repair outcomes by effectively guiding the model. \lnote{Furthermore, our empirical analysis demonstrates that the incompleteness of JPF itself has only a minor impact on the results, regardless of whether the context extractor is enabled or disabled.}
\end{resultbox}

\subsection{RQ4: The Quality of Generated Patches}

In this section, we discuss the repair quality of \ConcurGuard\ in comparison to the state-of-the-art automatic concurrency bug repair tools, \PFIX\ and \HIPPODROME. We quantify repair quality using the number of locks introduced during the repair process, a metric adopted from \PFIX~\cite{pfix}. 
This metric is motivated by the fact that excessive use of locks can lead to increased complexity, reduced concurrency, and a higher risk of deadlocks. 
This concern aligns with observations in the RacerD literature~\cite{RacerD}—the bug detection tool employed by \HIPPODROME—which notes that developers actively seek to minimize locking in practice, as superfluous locks inherently reduce program concurrency.

\begin{table}
  \centering
  \caption{The repair results of \ConcurGuard\ after five independent runs.}
  \label{tab:randomness}
  \begin{tabular}{cccccc}
    \toprule
       & $1$st  & $2$nd &$3$rd  & $4$th &$5$th  \\
    \midrule
    Correct Fixes          &29   &29     &28  &29   &27\\
    Avg Locks              &2    &2.17   &2.07&1.97 &2.21\\
    \bottomrule
  \end{tabular}
\end{table}

For all correctly repaired non-deadlock concurrency bugs, 
\PFIX\ introduces an average of 1.5 locks, 
while \ConcurGuard\ introduces 2 locks, 
and \HIPPODROME\ introduces 8.42 locks. 
\HIPPODROME\ introduces the most locks because it aims to protect 
all shared variables to eliminate data races. 
However, since only 8\%-10\% of data races are harmful~\cite{race-harm1, race-harm2}, 
this approach can be excessive. Additionally, \HIPPODROME\ sometimes introduces unnecessary protections due to existing happen-before relationships, which is an inevitable consequence of static analysis leading to false positives.
In contrast, \PFIX\ introduces the fewest locks on average 
because it is a dynamic analysis-based tool that targets bugs 
based on observed traces. 
Consequently, when \PFIX\ successfully fixes a bug, 
its repair is generally more precise and 
involves fewer additional locks.

In 5 of the 20 benchmarks where both \PFIX\ and \ConcurGuard\ successfully repaired bugs, \ConcurGuard\ introduced fewer locks, as it often favored atomic methods like \texttt{compareAndSet} over traditional locking. For details on patch quality from \ConcurGuard, please refer to our publicly available data.

\begin{resultbox}
    \textbf{Answer to RQ4}: 
Our experimental results show that \ConcurGuard\ introduces an average of 2 locks for correctly repaired non-deadlock concurrency bugs, significantly fewer than \HIPPODROME's 8.42 locks. While \ConcurGuard\ has more locks than \PFIX's average of 1.5, it avoids the excessive locking often seen in \HIPPODROME. Notably, in 5 out of 19 benchmarks where both tools successfully repaired bugs, \ConcurGuard\ used fewer locks, highlighting its ability to optimize lock usage and reduce complexity and deadlock risks.
\end{resultbox}

\section{Discussion}
\subsection{Randomness}

The repair results of \ConcurGuard\ exhibit a certain degree of randomness. Therefore, after independently running \ConcurGuard\ five times on non-deadlock concurrency bugs, we compared the total number of correctly fixed non-deadlock concurrency bugs with \HIPPODROME\ and \PFIX, as shown in Table~\ref{tab:randomness}. 

Using the Mann-Whitney U test, we found that \ConcurGuard\ achieved statistically significant improvements over both \HIPPODROME\ and \PFIX\ in repairing non-deadlock concurrency bugs, with a Mann-Whitney U statistic of 25 and a p-value of 0.0067. Furthermore, the number of locks correctly introduced in each run of \ConcurGuard\ was also analyzed, yielding a Mann-Whitney U statistic of 0 and a p-value of 0.0075 when compared to \HIPPODROME. These results indicate that \ConcurGuard\ significantly enhances repair quality compared to \HIPPODROME.

\subsection{Scalability}

As analyzed in Section~\ref{extractor}, the scalability of \ConcurGuard\ is not constrained by the context extractor. In practice, the extractor completes its analysis within 1 second for most benchmarks. To further evaluate its performance, we applied it to larger real-world programs used in prior work~\cite{Sword,D4}, including Sunflow (24,713 lines), Lusearch (48,128 lines), and Avrora (70,057 lines), achieving extraction times of 140s, 14s, and 77s, respectively.

Similar to \PFIX~\cite{pfix}, the primary scalability bottleneck of \ConcurGuard\ stems from state explosion—an issue that arises when detecting concurrency bugs via JPF~\cite{JPF}.
This challenge remains fundamentally difficult due to the exponential growth of the thread interleaving space, which is inherent to concurrency testing~\cite{explosion,pfix}.

\subsection{Oracle}
\lnote{First, we clarify the distinction between oracles based on bug causes and those based on bug symptoms~\cite{bug-survey}. 
Oracles grounded in bug causes—such as data races detected by tools like RacerD~\cite{RacerD}—provide information about specific types of interleavings (i.e., conflicting memory access pairs) that \textit{may} lead to incorrect behavior~\cite{race-harm1,hipp}.
In this setting, repair can be applied directly to the identified interleavings without requiring additional localization. By contrast, oracles based on observed bug symptoms—such as uncaught exceptions, assertion violations, or deadlocks during program execution—do not reveal the exact underlying interleaving that caused the symptom, necessitating localization before repair can be attempted. Consequently, repair in this case requires first locating the failure-inducing position corresponding to the observed symptom, making symptom-based approaches inherently more challenging~\cite{pfix,dyn-location}. \ConcurGuard\ falls squarely into this symptom-oriented category.}

\lnote{To evaluate our framework under different oracle settings, we extend our existing symptom-based evaluation with an additional cause-based setting. For cause-based oracles, following the methodology of \HIPPODROME, we target the elimination of data race warnings reported by RacerD. In this setting, the bug report component in our prompt is replaced with the race pairs identified by RacerD, effectively treating these warnings as the repair targets~\cite{RacerD}. This new evaluation demonstrates that our approach remains effective even when operating with cause-based oracles, complementing our original symptom-based assessment.}

\lnote{
On the non-deadlock concurrency bug dataset, \HIPPODROME\ achieves a 100\% correctness rate, consistent with its formal guarantees~\cite{hipp}, while \ConcurGuard\ attains 97.8\% when using GPT‑4, averaged over five runs per benchmark.
The lower correctness of \ConcurGuard\ stems primarily from its reliance on LLMs, whose inherently probabilistic nature makes it difficult to guarantee 100\% repair success under the same cause-based oracle setting. Manual inspection of the failed cases reveals that the LLM occasionally modifies program functionality in unintended ways, leading to incorrect repairs.}

\subsection{Soundness and Completeness of \ConcurGuard}

This section provides a systematic analysis of the soundness and completeness properties of \ConcurGuard.
Similar to other LLM-based repair methods~\cite{LLM-FIX1,LLM-FIX,LLM-FIX2,LLM-FIX3,response-format} and traditional program repair techniques~\cite{pfix,Grail}, \ConcurGuard\ does not offer theoretical guarantees of either soundness or completeness.

\lnote{Soundness refers to the property that every JPF-validated patch is indeed correct. Despite using JPF as a concurrency bug detector, our tool does not guarantee soundness. Using GPT-4 and executing each benchmark five times independently, we found that 5.4\% of patches passing JPF were actually incorrect when the context extractor was enabled, and 5.6\% when it was disabled (i.e., without source filtering). In both settings, the errors stemmed from two sources: JPF failing to trigger the bug (enabled: 2.3\%, disabled: 2.1\%), and the LLM unintentionally altering program semantics. The latter accounted for the remainder (enabled: 3.1\%, disabled: 3.5\%) and constituted a class of errors beyond JPF’s detection capability.}

Completeness denotes the extent to which our approach is able to fix bugs—that is, the proportion of bugs for which a correct patch can be produced. As reported in our evaluation (Section~\ref{sec:compare}), we measure empirical completeness by the fraction of bugs that are successfully repaired. Several factors influence the completeness of \ConcurGuard:

\lnote{First, the completeness of our approach is affected by JPF~\cite{JPF}, which is sound (no false positives) but not complete: due to the explosion of thread interleavings~\cite{explosion}, some bugs are missed. To assess the impact of JPF's incompleteness, we removed its bug reports from the benchmarks and observed only a slight decrease in repair accuracy.}

Second, completeness can be affected by the context extractor, as it may filter out code relevant to the bug. Although it lacks a theoretical completeness guarantee, results in Section~\ref{exep-extractor} show that no benchmark exists where the LLM succeeds without the extractor but fails with it. Hence, the extractor does not impair correctness and can improve repair accuracy in practice.

Third, the LLM inherently affects completeness. 
The first reason is that it can analyze only the source code itself, so bugs whose root causes lie outside the source code are beyond its analytical reach. This point has been clarified and quantified in Section~\ref{benchmark}. 
More importantly, the LLM itself does not provide guarantees of completeness.
To enhance the LLM’s repair capability, we provide precise contextual information and design improved prompting strategies, and we discuss their impact in detail in Section~\ref{exep-extractor} and Section~\ref{sec:strate}.

\subsection{Threats to Validity}

Internally, our main challenge is manually confirming the accuracy of the localization. Therefore, we use five people to independently judge whether the LLM has correctly localized the concurrency bug.
Another challenge is data contamination, as repairs for our benchmarks may already be present in the LLMs' training data. Based on our research on public datasets, these datasets typically only contain fixed code snippets rather than entire programs under test, which differs from our task. Additionally, we have not found any public fixes for some of the bugs in our benchmark. However, given the extensive and diverse nature of training data, completely eliminating contamination is difficult.
Lastly, due to the inherent opacity of LLMs, we cannot fully verify the correctness of certain explanations provided for SWE-agent and \ConcurGuard.
Externally, while our comprehensive study of 43 concurrent programs 
covers various types of concurrency bugs, 
our findings may not generalize to all concurrent programs. 
However, given our benchmark evaluation surpasses 
that of \PFIX\ and \HIPPODROME, 
we believe our results are indicative of performance 
across a broad range of scenarios.


\section{Related Work}

\subsection{APR for Concurrency Bugs}


\PFIX\ and \HIPPODROME\ represent the state-of-the-art in APR tools for concurrency bugs, utilizing dynamic and static analysis, respectively. \PFIX\ detects concurrency bugs through dynamic program execution and identifies their root causes via probabilistic statistics, then repairs them using predefined patch templates.
However, its probabilistic statistical approach may result in the actual bug root cause not being identified as the most suspicious~\cite{pfix}.
\HIPPODROME\ employs static analysis to detect data races and repairs this specific type of issues. However, it is limited to addressing only this category of issues and may fail to align with programmers’ intentions~\cite{oracle,oracle2}.
Our approach is most closely aligned with \PFIX, as both leverage dynamic analysis—
an approach that inherently eliminates false positives~\cite{oracle,oracle2}.
However, unlike \PFIX, which relies on statistical methods for bug localization and pre-defined repair templates, our method leverages the code comprehension capabilities of LLMs to locate and fix concurrency bugs.
Beyond \PFIX\ and \HIPPODROME, several other automated tools address concurrency bug repair. We categorize these tools based on the types of concurrency bugs they target.

There are various automated tools designed to repair different types 
of concurrency bugs. For example, some tools specifically target deadlocks. 
DFixer~\cite{deadlock1,deadlock2} is a recent tool that 
addresses deadlocks in C/C++ programs by preemptively acquiring locks, 
such as $w$ and $h$, to prevent the hold-and-wait condition.

On the other hand, several tools focus on fixing atomicity violations. 
AFix~\cite{AFix} leverages CTrigger's~\cite{CTrigger} output 
to add mutex locks but has been noted for potentially 
introducing deadlocks~\cite{Grail}. 
CFix~\cite{CFix} extends AFix’s approach to handle order violations, 
enforcing specific order relationships through synchronization methods. 
Similarly, AlphaFixer~\cite{CaiFix} targets atomicity violations 
in C++ programs by analyzing lock acquisitions to reduce deadlock risks, 
though there are concerns about its algorithm inadvertently 
introducing new deadlocks~\cite{CaiFix,hipp}.

There are also APR tools designed to address various types 
of concurrency bugs. Grail~\cite{Grail} claims to repair atomicity violations, 
data races, and non-communication deadlocks. 
However, it fails to account for related variables, 
which limits its ability to handle multi-variable bugs. 
Additionally, some patches generated by Grail can inadvertently introduce deadlocks~\cite{pfix}.
HFix~\cite{HFIX} employs syntactic or pattern-matching-based static analysis,
deriving patch patterns from a dataset of human-generated patches. 
This approach is constrained by the scope and quality of the human patches it considers, 
potentially limiting its effectiveness in diverse scenarios.

In contrast to these approaches, \ConcurGuard\ offers a more comprehensive solution. 
For example, while Grail struggles with multi-variable bugs and \PFIX\ 
does not address deadlocks, \ConcurGuard\ effectively handles atomicity violations, 
data races, and deadlocks. 
Moreover, \ConcurGuard\ is more efficient, as it neither relies on numerous bug and non-bug traces for probabilistic bug localization like \PFIX\ nor depends on constraint solvers for resolution, which can be overly burdensome.





\subsection{LLMs for APR}

The use of the LLMs in automated repair has been extensively explored, 
demonstrating their potential for effectively 
fixing bugs~\cite{LLM-FIX1,LLM-FIX,LLM-FIX2,LLM-FIX3,response-format}. 
Kolak et al.\cite{LLM-FIX3} utilized Codex along with two smaller LLMs 
to assess their ability to generate correct patches, 
highlighting the scalability of LLMs and the improved repair outcomes 
achievable with larger models. 
Xia et al.\cite{LLM-FIX} conducted a comprehensive study 
on the application of nine state-of-the-art pre-trained language models 
for APR across datasets from three different programming languages. 
Prenner et al.\cite{LLM-FIX1} performed a small-scale evaluation of 
the Codex model on a straightforward dataset containing Java and Python versions 
of buggy algorithm implementations. 
Xia et al.\cite{LLM-FIX2} tested newer LLMs on a broader range of benchmarks and 
compared their performance against multiple APR tools. 

Some researchers have also leveraged fine-tuned LLMs for APR, as fine-tuning enables models to learn from curated datasets of bug-fix pairs~\cite{Jiang2023Impact,fine-fix,RepairLLaMA,Yang2024Multi}. Jiang et al.~\cite{Jiang2023Impact} explored the impact of code language models on APR, demonstrating that fine-tuning can significantly enhance repair performance over naive prompting. Similarly, Huang et al.~\cite{fine-fix} conducted an empirical study on fine-tuning LLMs for repair, analyzing factors such as code representations and evaluation metrics. 
Silva et al. proposed RepairLLaMA~\cite{RepairLLaMA}, which applied parameter-efficient fine-tuning techniques as a key method in the field of program repair. This method substantially reduces computational overhead while maintaining competitive performance. Furthermore, Yang et al. proposed MORepair~\cite{Yang2024Multi}, a multi-objective fine-tuning framework that incorporates LLM-generated guidance to improve patch quality.

However, the aforementioned works lack the capability to perform end-to-end program repair—they treat bug fixing as an isolated process that still requires manual isolation of code snippets and provision of additional information. While some studies have developed end-to-end program repair tools~\cite{Dlfix,con-fix}, these cannot be applied to concurrent programs, as they fail to detect and localize concurrency bugs. Jin et al. addressed part of this gap by utilizing Infer for bug detection, localization, and contextual code extraction~\cite{Inferfix}, yet their approach is constrained to concurrency bugs arising solely from two-event interleavings. Thus, our work represents the first attempt to develop an end-to-end solution for the automated repair of diverse types of concurrency bugs using LLMs. In addition, recent advances have seen LLM-based agents deployed to automate complex tasks in digital environments~\cite{SWE,agentless,Autocoderover}. These approaches may leverage human interaction to support bug localization and repair, but our evaluation demonstrates that their effectiveness remains lower than that of our tool. Looking ahead, we plan to integrate such agent-based technologies with static analysis in future work, aiming to provide more accurate contextual information for further enhancing repair performance.


\section{Conclusion}

We introduce \ConcurGuard, an LLM-based agent designed to repair 
various types of concurrency bugs.
To our knowledge, it represents the first automated repair tool specifically designed for diverse concurrency bugs while enabling full-process automation, 
with no reliance on prior bug information or manual input.
Our experimental results demonstrate that leveraging LLMs for automated concurrency bug repair holds significant promise. Specifically, we find that the context extractor—which efficiently identifies code snippets relevant to concurrency bugs—plays a critical role in enhancing the LLM's repair capability. This finding underscores a key insight: providing LLMs with high-quality, relevant context is essential for unlocking their full potential. The more precise the relevant code provided to the LLM, the more effectively the model can analyze and address concurrency bugs.
Moving forward, we will explore more advanced techniques for concurrency bug-related context extraction to improve the accuracy of code snippets provided to LLMs and minimize irrelevant content.







\appendix

\begin{figure}[h!]
\centering\includegraphics[scale=0.83, trim=11.2cm 0.5cm 1.5cm 1.3cm, clip]{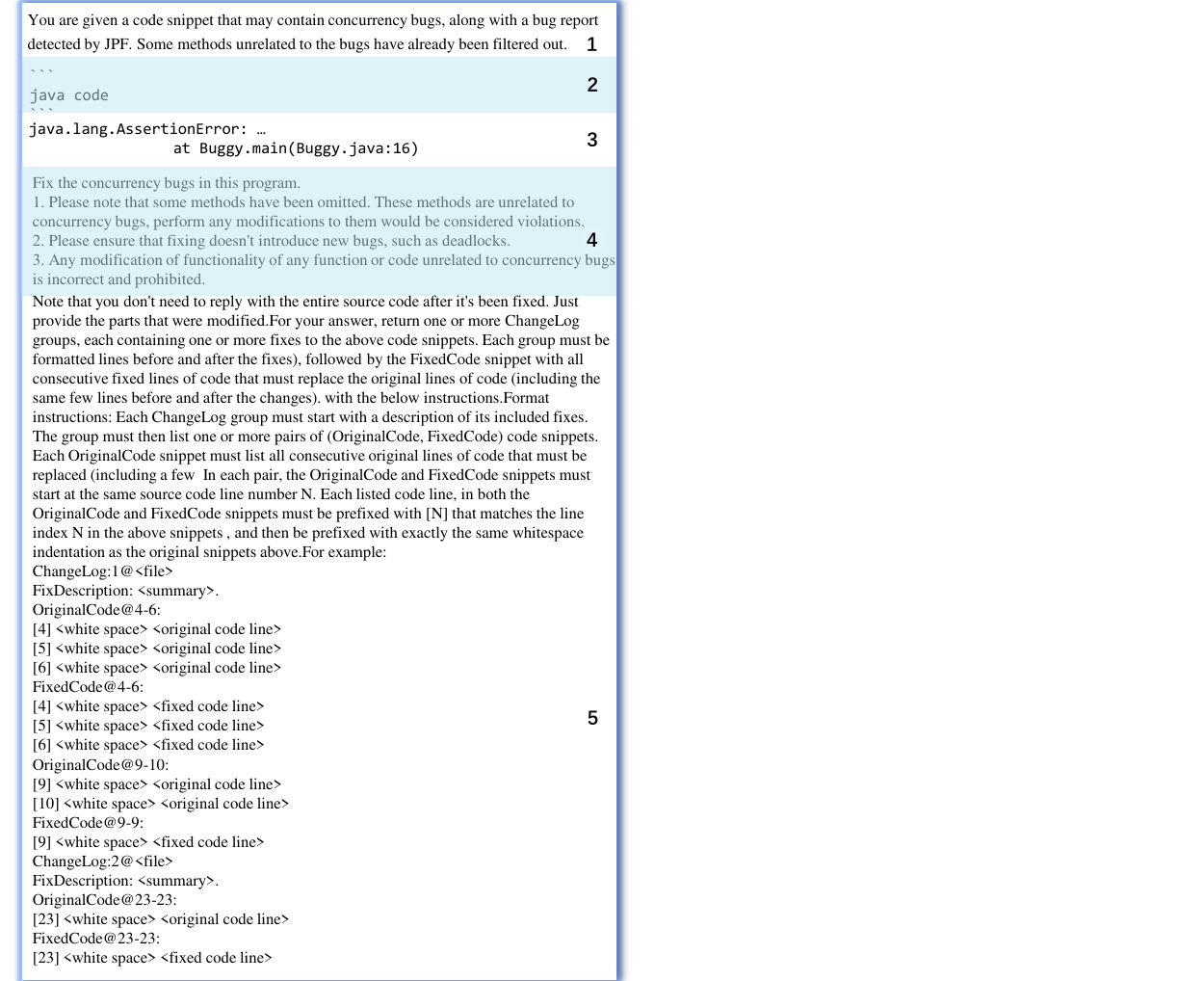}
\caption{The specific content of each part of the prompt for bug localization and fixing includes: Preclude, Code, Bug Report, Instructions, and Format Guidance.}
\label{fig:promptdetail}
\end{figure}

Figure~\ref{fig:promptdetail} presents the detailed content of each component in our prompt template for bug localization and fixing.

\end{document}